\documentclass[11pt]{article}
\usepackage{amsmath}
\usepackage{amssymb}
\usepackage{latexsym}
\usepackage{epsfig}

\newcommand{\RR}{\mathcal{R}}
\newcommand{\pp}{{\rm p}}
\newcommand{\XX}{{\rm X}}
\newcommand{\GG}{g}

\newcommand{\sectiono}[1]{\section{#1}\setcounter{equation}{0}}

\begin{document}

{}~
\hfill\vbox{\hbox{hep-th/0403051}\hbox{MIT-CTP-3476}
}\break

\vskip 3.8cm

\centerline{\Large \bf Twisted Tachyon Condensation
in Closed String Field Theory}

\vspace*{10.0ex}

\centerline{\large \rm Yuji Okawa and Barton Zwiebach}

\vspace*{8.0ex}

\centerline{\large \it Center for Theoretical Physics}

\centerline{\large \it
Massachusetts Institute of Technology}

\centerline{\large \it Cambridge,
MA 02139, USA}
\vspace*{1.0ex}

\centerline{okawa@lns.mit.edu, zwiebach@lns.mit.edu}

\vspace*{10.0ex}

\centerline{\bf Abstract}
\bigskip
\smallskip

We consider twisted tachyons
on $\mathbb{C}/\mathbb{Z}_N$ orbifolds
of bosonic closed string theory.
It has been conjectured that these tachyonic instabilities
correspond to decays of the orbifolds into flat space
or into orbifolds with smaller deficit angles.
We examine this conjecture using closed string field theory,
with the string field truncated to low-level tachyons.
We compute the tachyon potentials
for $\mathbb{C}/\mathbb{Z}_2$ and $\mathbb{C}/\mathbb{Z}_3$
orbifolds and find critical points at depths that
generate about 70\% of the expected change in the deficit angle.
We find that both twisted fields
and untwisted modes localized near the apex of the cone
acquire vacuum expectation values
and contribute to the potential.

\vfill \eject

\baselineskip=16pt

\tableofcontents

\sectiono{Introduction and brief summary} \label{s1}

In a stimulating  paper, Adams, Polchinski, and
Silverstein~\cite{Adams:2001sv}
gave a physical interpretation of the instability
associated with the twisted closed string tachyons
of the $\mathbb{C}/\mathbb{Z}_N$ orbifold
in Type II string theory.
In this orbifold two spatial dimensions form a cone
and twisted closed string states live
at the apex of the cone---an eight-dimensional subspace
of the ten-dimensional spacetime.
It was conjectured in \cite{Adams:2001sv} that
the instability signals
the possible decay of the cone into flat
two-dimensional space or into a cone with a smaller deficit angle.
Tests of this conjecture were discussed in detail
in~\cite{Adams:2001sv,Vafa:2001ra,Harvey:2001wm}
making use of probe D-branes
and the tools of conformal field theory (CFT).
Some aspects of the time-dependent decay process were examined
in the supergravity limit
\cite{Gregory:2003yb,Headrick:2003yu}. The tachyons themselves
do not feature in this analysis, which instead focuses on the radial
propagation of gravity and dilaton waves on the cone.  Presumably
these waves are generated by the condensation of twisted tachyons
at the apex.
Various issues concerning twisted closed string tachyons have been
discussed in~\cite{Dabholkar:1994ai,Dabholkar:2001gz,
David:2001vm,DeAlwis:2002kp}.

\smallskip
In this paper, we present a different approach
to twisted tachyon condensation.
We use closed string field theory (CSFT)
on the $\mathbb{C}/\mathbb{Z}_N$ backgrounds
to provide evidence for the conjecture in the framework of level
truncation.
The work of~\cite{Adams:2001sv} considered
Type II orbifolds $\mathbb{C}/\mathbb{Z}_N$ with odd $N$,
where the bulk tachyon is absent.
Since a covariant  closed superstring field theory
has not yet been constructed, we focus on bosonic strings, for
which  closed string field theory
exists~\cite{Zwiebach:1992ie,Saadi:tb,Kugo:1989tk,Kaku:zw}.
We believe that the interpretation for the instability
applies to the bosonic string,
despite the lack of supersymmetry
and the presence of a bulk tachyon.
Indeed, both the original orbifold $\mathbb{C}/\mathbb{Z}_N$
and a flat 2-space are consistent bosonic string backgrounds,
so it is natural to expect
that the twisted tachyons represent the instability of the
orbifold background to decay into flat space.
While the instability associated with the bulk tachyon
is still mysterious,
the bulk tachyon is present
both before {\em and} after the decay of the cone.
The instability we address is
the one associated with twisted tachyons.
A related situation occurs for the decay of
a wrapped D1-brane in bosonic string theory~\cite{Moeller:2000jy}.
The D1-brane has a spatially constant tachyon mode
and additional spatially dependent tachyon modes
that signal the instability to decay into a D0-brane.
When the D1-brane decays into a D0-brane,
the instability associated with the constant mode remains.

\medskip
The success in the computation of open string tachyon
potentials makes it natural to ask
if there is a closed string tachyon potential
that can be used to understand the decay of orbifold cones.
A conjecture concerning the critical points
of a twisted tachyon potential was put forward
by Dabholkar~\cite{Dabholkar:2001if}.
The conjecture relates the potential $\mathbb{V} (T)$
for twisted tachyons to the deficit angle $\theta$
of the conical orbifold geometry
via the relation $\theta = \kappa^2\, \mathbb{V} (T)$,
where $\kappa$ is the gravitational coupling constant.
For the $\mathbb{C}/\mathbb{Z}_N$ orbifold cone the deficit angle is
$\theta_N = 2\pi ( 1 - {1\over N})$.
In the string field theory formulated
around the orbifold background, the background itself is represented
by tachyons with  zero expectation values and the
tachyon potential $\mathbb{V}_N(T)$ satisfies
$\mathbb{V}_N(0) =0$.
If the endpoint of tachyon condensation is flat space,
the final deficit angle is zero.
If this background arises for tachyon expectation values $T_1$,
then $T_1$ must be a critical point of the potential
that satisfies
\begin{equation}
\label{theconjectureonpotential}
{ \kappa^2 \, \mathbb{V}_N(T_1)
\over 2\pi \bigl( 1 - {1\over N}\bigr) }
= -1 \,.
\end{equation}
This is the main prediction to be tested.
There are similar predictions
for critical points that represent transitions to orbifolds with
smaller deficit
angles. Unless noted otherwise, when we speak of the conjecture,
we refer to the expected relation
between deficit angles and critical points.

The above conjecture, however, needs refinement.
It turns out that expectation values for twisted fields
create nonvanishing one-point functions for bulk excitations
localized near the apex of the cone.
As a result, the physically relevant potential
depends on twisted degrees of freedom $\{ T \}$
and localized degrees of freedom
$\{ U_{\text{loc}} \}$ from
the untwisted sector:
\begin{equation}
\mathbb{V} = \mathbb{V} ( \{ T\} ;  \{ U_{\text{loc}}\} )\,.
\end{equation}
As opposed to conventional potentials $V(\phi)$ that are metric
independent and appear in
the Lagrangian as $\sqrt{-g}\, V(\phi)$, the potential $\mathbb{V}$
appears to have  metric dependence.
This dependence  arises because the expectation values for the
localized bulk modes are spatially dependent.  In addition, even
the couplings of twisted fields to the metric do not appear to be
conventional  since they contain metric dependent form factors.
The simple argument that relates
the potential of twisted fields
to the deficit angles (see \S\ref{section2.1})
is not strictly valid for a potential $\mathbb{V}$
with metric dependence,
so the conjecture~(\ref{theconjectureonpotential})
may not be strictly true.\footnote{We would like to thank M.~Headrick
for drawing our attention to the  metric dependence of the potential
and for useful discussions on its relevance
to the conjecture.}

Despite these caveats, we can frame our present analysis in terms of
the conjecture (\ref{theconjectureonpotential}).
In this paper, we truncate the string field
to tachyons whose levels are less than or equal to $1/4$,
a level far below the level of massless fields (level 2).
Since massless fields are not included,  the possible
effects of metric dependence do not feature in our present
computations.

\medskip
We were motivated to compute twisted tachyon potentials
in closed string field theory because, to date,
there is no evidence that critical points exist,
much less, that they have depths
proportional to the appropriate change of deficit angle.
The issue has been investigated in Type II superstrings
for the transition  $\mathbb{C}/\mathbb{Z}_N \to
\mathbb{C}/\mathbb{Z}_{N-2}$ in the limit of large $N$.
Here, according to the conjecture, the tachyon potential
should have a critical point at a depth of order $1/N^2$.
It has not yet been possible, however,
to confirm this expectation~\cite{Sarkar:2003dc},
although the question remains
under active investigation~\cite{Dabholkar,Adams}.
While these computations are done by naive
extrapolation from on-shell CFT results,
this may not be a problem since the tachyon
which is assumed to be responsible for the decay
is nearly massless.

In general, however,
the off-shell properties of string field theory
are crucial to obtain quantitative results.
In fact, to first approximation,
the truncation of the open string field to the tachyon
gave about 68\% of the D-brane energy~\cite{Sen:1999nx},
a rather encouraging result.
Had one used instead the on-shell tachyon action
to compute the D-brane energy, the answer
would have differed by a factor of $\RR^6$,
where $\RR = 3\sqrt{3}/4$ is a fundamental constant
in string field theory~\cite{Belopolsky:1994sk}.
The on-shell action would have predicted
about 329\% of the D-brane energy,
a rather discouraging result.
We will see that even larger off-shell factors arise
in closed string field theory.
Indeed, while CSFT calculations give
about 72\% of the expected depth
for the $\mathbb{C}/\mathbb{Z}_2$ orbifold,
the naive on-shell extrapolation would
predict about 1241\%
of the expected depth.

Twisted tachyons are expected to be simpler to
study than the bulk tachyon, for which no
translationally invariant vacuum has been
found~\cite{Belopolsky:1994sk,Belopolsky:1994bj}.
Twisted tachyons thus provide a new opportunity
to test the calculability of closed string field theory.
The viability of the CSFT level expansion is
supported by the remarkable fact that cubic interactions of
low level give results that are consistent
with the deficit angle predictions.
We find this encouraging
since all successful level-expansion computations
have followed this pattern: the key qualitative features
and the rough quantitative ones both emerge at low level.
We have also examined some
quartic interactions, and it appears that they do not
destabilize the results obtained with
the cubic interactions.
We emphasize that it is not yet clear what are the precise rules
of level expansion in closed string field theory. It is in fact
possible that interactions higher than cubic must be assigned an
intrinsic level.

In this paper we have worked
with the $\mathbb{C}/\mathbb{Z}_2$ orbifold,
which can only decay to flat space,
and with the $\mathbb{C}/\mathbb{Z}_3$ orbifold,
which can decay to flat space
and to the $\mathbb{C}/\mathbb{Z}_2$ orbifold.
In both cases,
the coupling of the twisted tachyons to localized excitations
of the bulk tachyon induce stabilizing quartic terms that
are {\em needed} to create critical points.
A rather rough analysis looking for
the flat space vacuum gives critical points
at about 96\% of the expected depth
for $\mathbb{C}/\mathbb{Z}_2$
and about 82\% of the expected depth
for $\mathbb{C}/\mathbb{Z}_3$.  With a more detailed
analysis that includes bulk tachyon self-interactions,
we find about 72\% of the expected depth
for $\mathbb{C}/\mathbb{Z}_2$
and about 99\% of the expected depth
for $\mathbb{C}/\mathbb{Z}_3$.
We also find the $\mathbb{C}/\mathbb{Z}_2$ vacuum
on the $\mathbb{C}/\mathbb{Z}_3$ tachyon potential
at about 134\% of the expected depth.
These results are very
encouraging but subject to some uncertainly.
As we discuss in some detail,
it seems possible to increase somewhat the level
of the bulk tachyon modes without having to include
new twisted fields and higher interactions.
If this is done,
the critical points become shallower and the predictions are
satisfied less accurately.
We expect that once new twisted fields
and higher interactions are included
the critical points will move downwards again.
The level expansion will not give a monotonic approach
to the expected results.
This is a  phenomenon familiar from
low-level lump computations~\cite{Moeller:2000jy}
and from high-level computations of the open string tachyon
vacuum~\cite{Gaiotto:2002wy,Taylor:2002fy}.
It is therefore important to perform higher-level computations
in the orbifold backgrounds.

\medskip
Technically, we need
two nontrivial pieces of
information. First, we need  the relation
between the closed string field
theory coupling constant and Newton's constant.
Second, we need the precisely normalized values
for the orbifold correlation functions
in the infinite volume limit.\footnote
{Some of the CFT computations in \S\ref{section3} have also
been performed independently in~\cite{Dabholkar}.}
Since our work depends crucially on using the correct numbers,
we give the detailed derivations of these results.

\sectiono{The conjecture and closed string field theory}
\label{lec1.1}

In the first part of this section we review
the conjecture put forward by Dabholkar~\cite{Dabholkar:2001if}
and discuss the refinements that seem necessary once
the unusual properties of the tachyon potential become evident.
After a discussion of these matters,
we briefly review closed string field theory
and state the precise relation
between the string field coupling constant and Newton's constant.

\subsection{The conjecture and its refinement}\label{section2.1}

The conjecture by Dabholkar~\cite{Dabholkar:2001if} is
motivated by the following effective field theory description
of closed strings in an orbifold background:
\begin{equation}
\label{motaction}
S =  \frac{1}{2 \kappa^2} \int d^Dx \sqrt{-g} e^{-2 \Phi} \, R
- \int_\mathcal{A} d^{D-2} x  \,
\sqrt{-g^{(D-2)}}\, e^{-2 \Phi} \, \mathbb{V}(T) \,.
\end{equation}
Here $\mathbb{V}(T)$ denotes the potential
for the twisted tachyon(s) $T$,  $\mathcal{A}$ is
the $(D-2)$ dimensional apex of the cone,
and $g^{(D-2)}$ is the determinant of the metric induced on
$\mathcal{A}$ by the full spacetime metric.
The twisted sector fields live on $\mathcal{A}$.

We look for backgrounds
where the dilaton $\Phi$ is constant over spacetime,
the metric describes the conical geometry
$\mathbb{R}^{D-2}\times
(\mathbb{C}/\mathbb{Z}_N)$, and the twisted tachyons
are constant over $\mathcal{A}$.   If the dilaton is constant,
its field equation requires
\begin{equation}
\label{dileqnfds}
\sqrt{-g} \, R   = (2\kappa^2) \, \mathbb{V} (T) \,
\,\sqrt{-g^{(D-2)}}\,\,\delta^{(2)} (x) \,.
\end{equation}
For the conical geometry   the metric factorizes into
a flat $(D-2)$ metric $g_{\alpha\beta} = \eta_{\alpha\beta}$
and a nontrivial two-dimensional metric $g_{ij}$.
For this spacetime metric
$R_{\alpha\beta}=0$, and $R$ is equal to the  scalar
curvature of the two-dimensional cone: $R=R^{(2)} = g^{ij} R_{ij}$.
Moreover $g = g^{(D-2)} g^{(2)}$,
where $g^{(2)}$ is the determinant of $g_{ij}$.
The above equation thus becomes
\begin{equation}
\label{conecurv}
\sqrt{g^{(2)}} \, R^{(2)} = (2\kappa^2) \, \mathbb{V} (T)
\,\,\delta^{(2)} (x) \,.
\end{equation}
Since the tachyons $T$ are constants on $\mathcal{A}$,
the above right-hand side is constant over $\mathcal{A}$,
as required by consistency given that the
left-hand side is also constant over $\mathcal{A}$.
For a cone of deficit angle
$\theta$ (or total angle $2\pi - \theta$), one has
$\sqrt{g^{(2)}}R^{(2)}= 2\theta \delta^{(2)} (x)$,
so (\ref{conecurv}) is
satisfied when the deficit angle and the potential are related by
\begin{equation}
\label{thetaVrel}
\theta = \kappa^2\, \mathbb{V} (T)\,.
\end{equation}
The gravitational field equations are also satisfied
by the conical background geometry.
The nontrivial components of Einstein's equations are
\begin{equation}
\sqrt{-g} \, R_{\alpha\beta} - {1\over 2} g_{\alpha\beta}
\Bigl( \sqrt{-g} \, R - 2\kappa^2 \,
\mathbb{V} (T) \, \sqrt{-g^{(D-2)}} \delta^{(2)}(x)
\Bigr) =0\,,
\quad R_{ij} - {1\over 2} g_{ij} R =0 \,.
\end{equation}
The dilaton equation (\ref{dileqnfds}) implies that
the object inside parentheses
in the first equation vanishes.
As a result, we find $R_{\alpha\beta}=0$,
which is satisfied by our metric.
With vanishing $R_{\alpha\beta}$, the second equation
is identically satisfied for a  two-dimensional cone.

For a transition between an initial configuration with deficit
angle $\theta_i$ and tachyon values $T_i$ and a final configuration
with deficit angle $\theta_f$ and tachyon values $T_f$,
equation (\ref{thetaVrel}) gives
\begin{equation}
\label{ybpred}
{\kappa^2\, (\mathbb{V} (T_i) -\mathbb{V} (T_f) )
\over \theta_i - \theta_f} = 1 \,.
\end{equation}
The deficit angle $\theta_N$ for the
$\mathbb{C}/\mathbb{Z}_N$ orbifold cone is
$\theta_N = 2\pi ( 1 - {1\over N})$.
Assume we begin with this orbifold background, represented
by tachyons with expectation values $T_N$. If the endpoint of
tachyon condensation is flat space, the final deficit angle is zero.
Let $T_1$ denote the expectation values of the tachyons
at a critical point that represents
flat space.
Equation (\ref{ybpred}) then predicts
that the following dimensionless ratio
must be equal to minus one:
\begin{equation}
{ \kappa^2 \, (\, \mathbb{V}(T_1) - \mathbb{V}(T_N)\, )
\over 2\pi \bigl( 1 - {1\over N}\bigr) }= -1 \,.
\label{f_N}
\end{equation}
For a transition from
the $\mathbb{C}/\mathbb{Z}_N$ orbifold
to the $\mathbb{C}/\mathbb{Z}_M$  orbifold,
with $2\leq M< N$, equation (\ref{ybpred}) predicts that
\begin{equation}
{ \kappa^2 \, (\, \mathbb{V}(T_M) - \mathbb{V}(T_N)\, )
\over 2\pi \bigl( {1\over M}- {1\over N}\bigr) }= \, -1 \,.
\label{f_NM}
\end{equation}
Here $T_M$  denotes the expectation values of the tachyons
at the critical point that represents
the $\mathbb{C}/\mathbb{Z}_M$ orbifold.
For any fixed $\mathbb{C}/\mathbb{Z}_N$ orbifold,
the above ratios provide $N-1$ tests of the calculated tachyon
potential. When we formulate the string field theory
of the  $\mathbb{C}/\mathbb{Z}_N$ orbifold, the tachyon
expectation values $T_N$ that represent the orbifold itself are
all zero. The tachyon potential, denoted henceforth as
$\mathbb{V}_N(T)$, automatically satisfies $\mathbb{V}_N
(T_N)=0$. It is therefore useful to define a normalized potential by
\begin{equation}
f_N (T) \equiv { \kappa^2 \,  \mathbb{V}_N(T)
\over 2\pi \bigl( 1- {1\over N}\bigr) }
\label{f_N-definition}
\end{equation}
so that the predictions in (\ref{f_N}) and (\ref{f_NM})
can be summarized as
\begin{equation}
f_N (T_M)
= \, - \,
{ \bigl( {1\over M}- {1\over N}\bigr)
\over \bigl( 1- {1\over N}\bigr) }  \,, \quad M= 1, 2, \ldots, N-1\,.
\end{equation}
We will study in detail the $\mathbb{C}/ \mathbb{Z}_2$ orbifold,
for which we have one test:
\begin{equation}
f_2(T) \equiv {1\over \pi  }\, \kappa^2 \, \mathbb{V}_2(T) \,, \qquad
f_2(T_1)\overset{?}{=} -1 \,.
\end{equation}
For the $\mathbb{C}/ \mathbb{Z}_3$ orbifold, which we also examine
in detail, we have two tests:
\begin{equation}
\label{z3predfac}
f_3(T) \equiv
{3\over 4\pi}\,\kappa^2 \,  \mathbb{V}_3(T) \,, \qquad
f_3(T_1) \overset{?}{=} -1\,, \quad
f_3(T_2) \overset{?}{=} -{1\over 4}\,.
\end{equation}
For any $\mathbb{C}/ \mathbb{Z}_N$ orbifold the values
$f_N(T_1), f_N(T_2) , \ldots\,,  f_N(T_{N-1})$
give the depths of the various critical points in the tachyon
potential, normalized using the expected depth of the point
that represents the transition to flat space.

The relation (\ref{thetaVrel}) between the deficit
angle and the potential $\mathbb{ V}$ is in fact familiar
in other guises. Indeed, for a cosmic string in four-dimensional
spacetime one has  $\theta = 8\pi G \mu$, where $8\pi G = \kappa^2$
and $\mu$ is the energy per unit length
of the string~\cite{Vilenkin:zs,Vilenkin:ib}.
In three-dimensional gravity, one has $\theta = 8\pi G M$,
where $M$ is the mass of a point particle that creates a conical
geometry with deficit angle $\theta$~\cite{Deser:tn}.  The energy
associated with $\mathbb{V}$,
or its lower dimensional versions
for strings or point particles,
may have an intrinsic gravitational meaning.
It has been demonstrated that
spaces whose asymptotic geometry is conical
can be assigned an energy
through a comparison with asymptotically flat
space~\cite{Hawking:1995fd}.
This energy agrees with value of the potential $\mathbb{V}$.

\medskip
The explicit analysis in the following sections
demonstrates that the above conjecture needs refinement.
The relevant potential is not just a potential
for all the scalars in the twisted sector.
Expectation values for twisted
fields create nonvanishing one-point functions for
modes of bulk fields that live near the apex of the cone.
It follows that these modes must be excited and, as we shall
see, they contribute to the potential.
The bulk tachyon, for example, acquires some expectation
value---roughly a Gaussian that
decays quickly away from the apex of the cone.
We claim that the physically relevant potential depends
on twisted fields {\em and} localized degrees of freedom from
the untwisted sector:
\begin{equation}
\label{stypepot}
\mathbb{V} = \mathbb{V} ( \{ T\} ;  \{ U_{\text{loc}}\} )\,,
\end{equation}
where $\{ T\}$ denotes the infinite collection of twisted
fields (thus the $T$), and $\{ U_{\text{loc}}\}$ denotes
the infinite collection of localized excitations
in the untwisted sector (thus the $U$).
At least for bosonic strings,
the existence of a critical point seems to
require both kinds of excitations.

In general,  the change of deficit
angle in the process of tachyon condensation
cannot be read directly from a potential of the type
indicated in (\ref{stypepot}).
To explain this consider
a scalar $\phi$ minimally coupled to pure gravity,
with a scalar potential $V(\phi)$ that, as usual,
is metric independent.
The problem of finding a
vacuum is easily solved in two steps.
One finds the critical points of the potential $V(\phi)$ and sets
$\phi$ equal to any of the critical values.
The potential evaluated at the chosen critical point defines the
cosmological term.
The gravitational equations can then be solved
to find the appropriate de~Sitter
or anti-de~Sitter background.
Similar remarks apply to the idealized system described by
the action~(\ref{motaction}):  the values of the
potential are directly correlated to the observable
deficit angle of the cone.  The  problem arises for
the potential (\ref{stypepot}) because the
contributions from the localized
bulk fields depend nontrivially on the
spacetime metric; the corresponding
expectation values are obtained
by solving equations of motion
with metric dependent kinetic terms.
The (unusual) metric dependence
of the potential $\mathbb{V}$ implies that its minimization cannot be
disentangled from the problem of solving the gravitational equations.
Even if we calculated $\mathbb{V}$ to full accuracy using the
fixed metric of the original orbifold,
the critical values of $\mathbb{V}$ need not reproduce
the precise expectations for the deficit angles.
Indeed, at any critical point
where a new orbifold would be expected,
the metric is different from the original one,
and the potential is not strictly applicable.

In computations of $\mathbb{V}$
that only include fields with levels lower than two,
one is in effect using the background metric
of the original orbifold.
Since the massless closed string fields appear at level two,
neither metric changes nor gravitational back-reaction feature
in the low-level calculations that we perform in the present paper.
The issue of metric dependence does not arise and one is
then justified in relating the depth of critical points in
$\mathbb{V}$ to changes in deficit angles.
This is what we do in the sections that follow.

\subsection{Closed string field theory and Newton's constant}

In this section we briefly recall
the structure of the closed string field theory action
and give the computational rules
for the quadratic and cubic terms.
Some discussion of the  quartic couplings
will be given in \S\ref{elem4pt}.   While most of our
results will be general, the
theory around flat space will receive particular attention.
The formulation of CSFT for the orbifold background
will be discussed in the later sections.

The closed string field theory action around flat space
is given in~\cite{Zwiebach:1992ie}:
\begin{equation}
\label{action}
S = -{2\over \alpha'} \Bigl(~ {1\over 2}
\langle \Psi\,, c_0^- Q \Psi\rangle
+ {1\over 3!} \,  \kappa\,  \langle \Psi, \Psi, \Psi\rangle
+ {1\over 4!} \,  \kappa^2\,  \langle \Psi, \Psi, \Psi, \Psi\rangle
+ \, \ldots \,\Bigr)~~\,,
\end{equation}
where the dots represent additional
interactions---the theory is fully nonpolynomial.
The closed string field $|\Psi\rangle$ is a ghost number
two vector of the CFT state space and satisfies the constraints
$(b_0 - \bar b_0) |\Psi\rangle = (L_0 - \bar L_0) |\Psi\rangle=0$.
The above action, as opposed to that in~\cite{Zwiebach:1992ie},
includes the explicit $\alpha'$ dependence.
We use a spacetime metric with signature
$(-, +, + , \ldots , + )$
and $c_0^\pm = {1\over 2} ( c_0 \pm \bar c_0)$.
The BRST operator $Q$ is normalized by writing
$Q = c_0L_0 + \bar c_0 \bar L_0 + \ldots =
c_0^+ (L_0 + \bar L_0) + c_0^- (L_0 - \bar L_0) + \ldots\,,$
where the dots indicate terms that involve
neither $c_0$ nor $\bar c_0$.
For tachyon states $c_1 \bar c_1 |p\rangle$ of momentum $p$
we have $L_0 = \bar L_0 = -1 + {1\over 4} \alpha' p^2$.
The momentum operator $\widehat p$ is Hermitian:
$\widehat{p}^{\, \dagger} = \widehat{p}$,
but under BPZ conjugation
$\widehat p\to - \widehat p$.
This  means that Hermitian conjugation takes
$ |p\rangle \to \langle p|$
while BPZ conjugation takes $ |p\rangle \to \langle -p|$.
The BRST operator is Hermitian, and the above action is real
if the Hermitian conjugate of the string field $\Psi$ is
equal to {\em minus} the BPZ conjugate of the field.

We define the inner product $\langle \cdot \,, \cdot \rangle$
in terms of BPZ conjugation and an overlap between bras and kets:
$\langle A, B\rangle = \langle bpz(A) | B\rangle$.
In the theory formulated around flat spacetime,
the basic overlap is
\begin{equation}
\langle p' | \, c_{-1} \bar c_{-1} \,\, c_0^- c_0^+ \,\,
c_1 \bar c_1 | p\rangle = (2\pi)^D \, \delta^{(D)} ( p- p') \,.
\end{equation}
With ghost fields $c(z) = \sum_n c_n z^{-n+1}$ and $\bar c (\bar z)
= \sum_n \bar c_n {\bar z}^{-n+1}$ one has
\begin{equation}
\langle p' | \, c(z_1)\bar c(\bar z_1) \, c(z_2)\bar c(\bar z_2) \,
c(z_3)\bar c(\bar z_3) \, | p\rangle
= 2\, (2\pi)^D \, \delta^{(D)} ( p- p') |z_1 - z_2|^2 |z_1 -
z_3|^2 |z_2 - z_3|^2\,.
\end{equation}

For the purposes of the present paper
it suffices to consider string fields of the form
$|\Psi\rangle = \sum_\alpha  c_1 \bar c_1  \phi^\alpha
|\mathcal{O}_\alpha\rangle$, where $\alpha$ is
an index that can take both continuous and
discrete values, $\phi^\alpha$ is a component field,
and $|\mathcal{O}_\alpha\rangle$ is a {\em primary} state
of the matter conformal field theory with conformal dimensions
$h_\alpha = \bar h_\alpha$.
For this string field, the quadratic and cubic terms
in the action are
\begin{equation}
\label{evalS3}
S =-{2\over \alpha'} \sum_{\alpha, \beta}  (h_\beta -1) \,
\phi^\alpha\, m_{\alpha\beta}\, \phi^\beta
-{1\over 3!} {2\kappa\over \alpha'} \sum_{\alpha, \beta, \gamma}
\,  \RR^{6-2(h_\alpha+ h_\beta+
h_\gamma)} \,\,\phi^\alpha \phi^\beta \phi^\gamma \,
\mathcal{C}_{\alpha\beta\gamma} \,.
\end{equation}
Here $m_{\alpha\beta}
= \langle bpz (\mathcal{O}_\alpha)| c_{-1} \bar c_{-1} \,
c_0^- c_0^+ \, c_1 \bar c_1 | \mathcal{O}_\beta\rangle$
and the constant $\mathcal{C}_{\alpha\beta\gamma}$ is defined
by the CFT correlator
\begin{equation}
\bigl\langle c \bar c \mathcal{O}_\alpha (0) \, c \bar c
\mathcal{O}_\beta (1)\,  c \bar c \mathcal{O}_\gamma (\infty)
\bigr\rangle  = \mathcal{C}_{\alpha\beta\gamma}\,.
\end{equation}
This correlator is evaluated in the complex plane $z$,
with local coordinates $z,  z-1$, and $1/z$ around the points
$0,1$, and $\infty$, respectively.
The prescription (\ref{evalS3}) for the evaluation
of the cubic term follows from the geometrical description
of the three-closed-string vertex.
The vertex can be viewed as the Riemann sphere
$\widehat{\mathbb{C}}$ punctured at
$1/\sqrt{3}, \omega /\sqrt{3},$ and $\omega^2/\sqrt{3}$,
where $\omega= \exp (2\pi i/3)$.
The distance between any two punctures is one.
The radial lines departing $z=0$ at angles
$\pm\pi/3$ and $-\pi$ cut the sphere into three punctured disks.
The constant $\RR$ is the inverse of the mapping radius\footnote
{The mapping radius $\rho$ of a disk
$D$ with coordinate $z$ at the puncture is given by $|dz/d\xi|$,
evaluated at the puncture,
where $z(\xi)$ is the conformal map from a unit disk
$|\xi| =1$ punctured at the origin to $D$.
See \cite{Belopolsky:1994sk} for more details.}
$\rho$ of any of these three punctured disks:
\begin{equation}
\label{maprad}
\RR \equiv {1\over \rho}=  {3\sqrt{3}\over 4} \simeq
1.2990\,.~
\end{equation}
The constant $\RR$ plays an important role
in any off-shell amplitude.
Indeed, the $\RR$ factor in the cubic term
is the product of factors of the form $\RR^{2(1-h)}$,
one for each operator $c_1 \bar c_1 \mathcal{O}$.
For an on-shell operator, the factor becomes unity since $h=1$.
For off-shell operators, however, the factors of $\RR$
can have a significant effect.

Consider now the CSFT action in $D=26$,
evaluated for the tachyon string field:
\begin{equation}
| \Psi \rangle
= \int {d^Dp\over (2\pi)^D}  \, u(p) \, c_1 \bar c_1
|p\rangle\,, \qquad   u(p) =
\int d^Dx  \, u(x) e^{-i p \cdot x}\,, \quad u^*(p) = u(-p)\,.
\label{bulk-tachyon-string-field}
\end{equation}
Using the rules described above,  the quadratic and
cubic terms in the action take the form
\begin{equation}
\label{utachefffac}
-{1\over 2} \int\hskip-5pt  {d^Dp\over (2\pi)^D}   u(-p)
\Bigl( p^2 - {4\over
\alpha'}\Bigr)  u(p) -{1\over 3!}   {4\kappa\over \alpha'} \,
\int \prod_{i=1}^3 \Bigl[ {d^Dp_i\over (2\pi)^D}
\,\RR^{2 - {1\over 2} \alpha'  p_i^2} \, u(p_i)\Bigr]
(2\pi)^D \,\delta^{(D)} (p_1+p_2+p_3)\,.
\end{equation}
In coordinate space this tachyon action $S_u$ becomes
\begin{equation}
\label{tachactionbulk}
S_u = \int d^Dx
\Bigl\{ -{1\over 2}\, \eta^{\mu\nu} \partial_\mu u
\partial_\nu u  -{1\over 2}\Bigl( -{4\over
\alpha'}\Bigr) u^2 -{1\over 3!}{4\kappa\over \alpha'} \,
\Bigl[ \RR^{2 +{1\over 2} \alpha' \partial^2 }
u \Bigr]^3 \Bigr\} \,.
\end{equation}
The S-matrix element for the scattering
of three on-shell tachyons is therefore
\begin{equation}
\label{sft3pt}
\mathcal{S} = -i  \,{4\kappa\over \alpha'} \,
(2\pi)^D \delta^{(D)} (\sum_{i=1}^3 p_i)\,.
\end{equation}
This result can be compared with the familiar CFT
calculation for this amplitude
in terms of the conventional gravitational coupling constant
(see \cite{polch}, equations (6.6.9) and (6.6.18), and note the
difference in overall sign for the tachyon field).
One finds that the coupling $\kappa$ in the CSFT action
{\em is} the gravitational coupling constant.\footnote
{In four spacetime dimensions $\kappa^2 = 8 \pi G$,
where $G$ is Newton's constant.}
The spacetime  gravitational action
that emerges from the closed string field theory is therefore
given by the  first term in (\ref{motaction}).

The tachyon potential $\mathbb{V}(u)$ is obtained from $S_u$
by taking all spacetime derivatives to vanish:
\begin{equation}
\mathbb{V}(u) = -{1\over 2}\Bigl( {4\over
\alpha'}\Bigr) u^2 + {1\over 3!}{4\kappa\over \alpha'} \,
\RR^6\, u^3 \,.
\end{equation}
This potential has a nonvanishing critical point $u_*$ and
\begin{equation}
\label{predcosm}
\mathbb{V}(u_*) = {1\over 2\kappa^2} \, {1\over \alpha'}
\Bigl( - {16\over 3 \RR^{12}} \Bigr)
\simeq {1\over 2\kappa^2} \, {1\over \alpha'} (-0.23096) \,.
\end{equation}
The meaning of this critical point, first found
in~\cite{Kostelecky:1990mi}, is not
clear.   The critical point in fact disappears
upon inclusion of the quartic term in the tachyon
potential~\cite{Belopolsky:1994sk,Belopolsky:1994bj}.\footnote
{It has been suggested by Bergman~\cite{bergman}
that $\mathbb{V}(u_*)$ represents
a cosmological constant that may be compared with the
noncritical string cosmological term
$\Lambda = -{1\over 2\kappa^2}{1\over \alpha'}
{2(D-26)\over 3}$~(see~\cite{polch}, equation (3.7.20))
for $D\leq 2$.
This value of $\Lambda$ for $D=2$, however,
is approximately 70 times
larger than the value of $\mathbb{V}(u_*)$.
This comparison, if valid, probably requires a re-computation
of the lower-dimensional gravitational coupling.}

\medskip
In open string field theory (OSFT)
level expansion has proven
to be a surprisingly good approximation scheme.
The level $\ell$ of a state
is defined to be equal to the $L_0$ eigenvalue of the state plus one.
As a result, the zero-momentum open string tachyon has level zero.
The level of the spacetime field associated
with a state is declared to be equal to the level of the state.
For closed strings we define
\begin{equation}
\label{cslevel}
\ell \equiv L_0 + \overline L_0 + 2 \,.
\end{equation}
It follows that the level of the  primary state
$c_1 \bar c_1 |\mathcal{O}_\alpha\rangle$  is
\begin{equation}
\label{cstaprimsdnlevvel}
\ell \, \bigl( c_1 \bar c_1 |\mathcal{O}_\alpha\rangle \bigr)
= 2h_\alpha\,.
\end{equation}
For example,  the level of the closed string tachyon state
$c_1 \bar c_1 |p\rangle$ with momentum $p$ is
\begin{equation}
\label{cstachyonlevvel}
\ell \, \bigl( c_1 \bar c_1 |p\rangle \bigr)
= {1\over 2} \, \alpha' p^2\,.
\end{equation}
In this convention, the level of a massless closed string state
is equal to two.
The level of a cubic closed string interaction
is defined as in OSFT:
it is equal to the sum of levels of the operators
that appear in the interaction.
The level of the operators does in fact enter into the
explicit form of the cubic term.  For primary fields
a cubic interaction of total level $L$ contains a suppression
factor $1/\RR^L$, as can be seen from (\ref{evalS3}).
This suppression is presumably one of the various ingredients
that make level expansion work.
We suspect that quartic and higher elementary interactions
in CSFT carry intrinsic level.
This would mean that the total level of any such interaction would
be given by the sum of levels of the operators
plus the intrinsic level carried by the interaction.
The intrinsic level, for example, may be given by a constant times
$(k-3)$, where $k$ is the order of the interaction.
If this turns out to be the case, finite level computations
will only include a finite set of interactions
despite the nonpolynomiality of the theory.

\sectiono{Orbifold CFT correlators}\label{section3}

In this section we compute correlators of the orbifold
CFT~\cite{Dixon:1986qv,Hamidi:1986vh,Bershadsky:nh}
which are necessary
for our computations of the tachyon potentials. Much of the
needed work can be taken directly from the detailed work of
Dixon {\em et~al.}~\cite{Dixon:1986qv}.
Since precise numerical values are crucial in our work,
we introduce the necessary factors of $\alpha'$
in all of the formulae---this also helps keep track of units.
Moreover, we must take the infinite volume limit
of the results in~\cite{Dixon:1986qv},
which are given for two-dimensional orbifolds
with finite volume $V_\Lambda$.\footnote
{Some aspects of conformal field theory
for noncompact spacetimes were discussed in \cite{Kraus:2002cb}.}
Correlators of three twist fields,  for example,
are proportional to $\sqrt{V_\Lambda}$ and appear
to become infinite when  $V_\Lambda\to \infty$, but proper
attention to the normalization of the operators involved shows
that effectively the volume $V_\Lambda$ is replaced
by a finite constant.

\smallskip
Consider the two dimensional plane spanned by the
string coordinates $X^1$ and $X^2$.  This plane can be
described as the complex
plane $\mathbb{C}$  with a complex coordinate $\XX$:
\begin{equation}
{\XX} = X^1 + i X^2, \quad \bar{\XX} = X^1 - i X^2.
\end{equation}
The orbifold $\mathbb{C}/\mathbb{Z}_N$
is obtained by $\mathbb{Z}_N$
action on the complex coordinate $\XX$.
The generator of this action takes
$\XX$ to $e^{2\pi i/N} \XX$.
The quotient space is a cone (orbifold) whose total angle
at the apex is $2\pi/N$; the deficit angle is
$2\pi (1 - {1\over N})$.
It is a noncompact orbifold of infinite volume. In the
orbifold CFT there are twist fields $\sigma_k(z,\bar z)$
labelled by the integer
$k$, with $0<k<N$.
If we place $\sigma_k$ at $z=0$, the coordinate field has
the monodromy
$\XX (e^{2\pi i} z, e^{-2\pi i} \bar z) = e^{2\pi i k/N}
\XX (z, \bar z)$.
The conformal dimensions of $\sigma_k$ are
\begin{equation}
\label{dimtwist}
h_k=\bar {h}_k={1\over 2}\,\frac{k}{ N} \left( 1 - \frac{k}{N}
\right)\,.
\end{equation}
The relevant CFT is the full CFT obtained as the tensor product
of the CFT which describes the flat $D-2$ dimensions and the
orbifold CFT of the cone (the inclusion of the ghost CFT
causes no significant change).  The twist
fields are normalized so that two-point functions
on the sphere are given~by
\begin{equation}
\label{basic2ptfunction}
\langle \sigma_{N-k} (\infty) \, \sigma_k (0) \, \rangle
= V_{D-2}\,.
\end{equation}
The insertion at $z=\infty$, as usual, is defined by an insertion
at $w=0$, with $w=1/z$.  When the subscripts of the  twist fields
do not add up to $N$ the correlator vanishes.
The factor $V_{D-2}$ denotes the volume
of the flat lower-dimensional spacetime.
This factor is necessary because the twist operators
can be dressed with momenta $q$ in the $D-2$ directions
in which case
\begin{equation}
\label{basicilg2ptfunction}
\langle  e^{iq\cdot X}\sigma_{N-k} (\infty) \,  e^{iq'\cdot
X}\sigma_k (0) \, \rangle =
(2\pi)^{D-2} \delta^{(D-2)} (q+ q')\,.
\end{equation}

It is customary to  fix  some $k$ that satisfies  $0< k \le N/2$
and to call $\sigma_+ \equiv\sigma_k$.
Additionally, we let $\sigma_- \equiv \sigma_{N-k}$,
which is reasonable since the monodromies around $\sigma_-$
and $\sigma_+$ are inverses of each other.
The twist fields $\sigma_{+}$ and $\sigma_{-}$ have
identical dimensions, given by (\ref{dimtwist}).
Finally, one introduces `doubly'
twisted fields:
$\sigma_{++} \equiv \sigma_{2k}$
and $\sigma_{--} \equiv \sigma_{N-2k}$ for $0< k < N/2$,
which also have identical dimensions.
On account of (\ref{basic2ptfunction}),
$\langle \sigma_-(\infty) \, \sigma_+(0)\rangle
= \langle \sigma_{- -}(\infty) \, \sigma_{+ +}(0)\rangle =V_{D-2}$.

\medskip
The orbifold CFT admits `untwisted' operators that carry momentum
along the directions on the cone.  These operators are needed
to represent bulk fields that live everywhere,
including the apex of the cone.
Using the momenta $p_1$ and $p_2$
associated with $X^1$ and $X^2$, respectively,
we introduce the complex combinations:
\begin{equation}
\pp = \frac{1}{2} ( p_1 - i p_2 ) , \quad
\bar \pp = \frac{1}{2} ( p_1 + i p_2 )\quad \to \quad
p_1 X^1 + p_2 X^2 =  \pp \XX + \bar{\pp} \bar {\XX} \,.
\end{equation}
We also define  $p^2 \equiv p_1^2 + p_2^2 = 4 \,\bar{\pp} \pp$.
The  vertex operator
\begin{equation}
{\cal U}_\pp \equiv  e^{\, i {\rm p X} + i {\rm \bar{p} \bar{X}}},
\end{equation}
does not  belong to the orbifold CFT since it is not
invariant under $\pp \to \alpha \,\pp$
with $\alpha = \exp (2\pi i /N)$.
An invariant vertex operator ${\cal V}_\pp$ is readily defined:
\begin{equation}
{\cal V}_\pp \equiv
\frac{1}{N} \sum_{i=0}^{N-1} {\cal U}_{\alpha^i \pp}
= \frac{1}{N} \, ( {\cal U}_{\pp} + {\cal U}_{\alpha \pp} + \cdots
+ {\cal U}_{\alpha^{N-1} \pp} ) \quad\to\quad
{\cal V}_{\alpha\,\pp} = {\cal V}_\pp\,.
\end{equation}
A complete set of independent operators $ {\cal V}_\pp$
can be chosen by restricting $\pp$
to any fundamental domain $\Gamma$
of the identification $\pp \sim \alpha \,\pp$.
The operators ${\cal U}_\pp$ (and ${\cal V}_\pp$)
have dimensions $ h = \bar{h} ={1\over 4} \alpha' p^2 = \alpha'
\bar{\pp}\, \pp$ and are normalized such that
\begin{equation}
\label{standucorr}
\langle {\cal U}_\pp (\infty) \, {\cal U}_{\pp'} (0) \rangle
=  (2 \pi)^2 \delta^{(2)} (p+p') \, V_{D-2}\equiv
(2 \pi)^2 \delta (p_1+p_1') \delta (p_2+p_2') \, V_{D-2}\,.
\end{equation}
This equation must be viewed as a computational device to obtain
two-point functions of well-defined operators of the orbifold CFT.
Since ${\cal U}_0 = {\bf 1}$, we find
\begin{equation}
\langle\, {\cal U}_\pp (0) \rangle = (2 \pi)^2 \delta^{(2)} (p) \,
V_{D-2}\,.
\end{equation}

\medskip
We can use the above discussion to produce the  vertex operator for
a general configuration of the bulk
tachyon.
This is given by a superposition of states of various momenta
$q$, $p_1$, and $p_2$:
\begin{equation}
\int \frac{d^{D-2} q}{(2 \pi)^{D-2}}
\int_\Gamma \frac{d^2 p}{(2 \pi)^2} \,
\,N  u(q, \pp) \, c \bar{c} \, e^{i q \cdot X} \, {\cal V}_\pp\,,
\end{equation}
where we included a factor of $N$ and integrated over the fundamental
domain $\Gamma$, with $d^2p\equiv dp_1 dp_2$.
The weight function $u(q,\pp)$ will be eventually
identified as the bulk tachyon field.
Using the definition of ${\cal V}_\pp$ we find
\begin{equation}
\int \frac{d^{D-2} q}{(2 \pi)^{D-2}}
\int_\Gamma \frac{d^2 p}{(2 \pi)^2} \,
u(q,\pp) \, c \bar{c} \, e^{i q \cdot X} \,( {\cal U}_{\pp}
+ {\cal U}_{\alpha \,\pp} + \cdots
+ {\cal U}_{\alpha^{N-1} \pp} )\,.
\end{equation}
It is convenient to extend the definition of the function $u(q,\pp)$
by setting
$u(q, \alpha \,\pp) = u(q,\pp)$.
This definition allows us to rewrite the vertex operator as
\begin{equation}\text{bulk-tachyon vertex operator:}\quad
\int \frac{d^{D-2} q}{(2 \pi)^{D-2}}
\int \frac{d^2 p}{(2 \pi)^2} \,
u(q,p) \, c \bar{c} \, e^{i q \cdot X} \, {\cal U}_\pp\,.
\label{bulk-tachyon-vertex-operator}
\end{equation}
Here the two-dimensional integration is unrestricted  and, for
simplicity of notation,
we write $u(q,p)$ instead of $u(q,\pp)$.
This vertex operator will be of
utility in \S\ref{secz2pot} and \S\ref{lec1.2}.

\medskip
Let us now discuss operator product coefficients and three-point
functions that involve twist fields.
We introduce
coefficients
$C^{\cal V}_{-,+} (\pp)$, with $\pp \in \Gamma$,
using the operator product expansion (OPE) of
$\sigma_{-}$ and $\sigma_{+}$:
\begin{equation}
\sigma_{-} ( x, \bar{x} ) \, \sigma_{+} (0)
\sim \int_\Gamma \frac{d^2 p }{(2 \pi)^2} \,N\,
C^{\cal V}_{-,+} (\pp) \, |x|^{2 \alpha' \bar{\pp} \pp
- \frac{2k}{N} \left( 1 - \frac{k}{N} \right) } \,
{\cal V}_\pp (0) +\ldots \,,
\label{sigma-sigma-OPE}
\end{equation}
where the dots denote terms with operators
that have vanishing one-point functions
and vanishing two-point functions with ${\cal V}_\pp$.
If we define $C^{\cal V}_{-,+} (\pp)$ for any $\pp$ by requiring
$ C^{\cal V}_{-,+} (\alpha\, \pp) = C^{\cal V}_{-,+} (\pp)$,
then (\ref{sigma-sigma-OPE}) can be rewritten as follows:
\begin{equation}
\label{sigma-sigma-OPE-2}
\sigma_{-} ( x, \bar{x} ) \, \sigma_{+} (0)
\sim \int \frac{d^2 p }{(2 \pi)^2} \,
C^{\cal V}_{-,+} (\pp) \, |x|^{2 \alpha' \bar{\pp} \pp
- \frac{2k}{N} \left( 1 - \frac{k}{N} \right) } \,
{\cal U}_\pp (0) +\ldots
\end{equation}
It follows from this OPE that
\begin{equation}
\langle \sigma_{-} (1) \, \sigma_{+} (0) \rangle
= \int \frac{d^2 p }{(2 \pi)^2} \,
C^{\cal V}_{-,+} (\pp) \, \langle {\cal U}_\pp (0) \rangle
= C^{\cal V}_{-,+} (0)\,  V_{D-2}\,.
\end{equation}
Therefore,
\begin{equation}
\label{normatzeromom}
C^{\cal V}_{-,+} (0) = 1\, .
\end{equation}
Let us define the three-point function $C_{{\cal V},-,+} (\pp)$ by
\begin{equation}
C_{{\cal V},-,+} (\pp) = \langle  {\cal V}_\pp (\infty) \,
\sigma_{-} (1) \, \sigma_{+} (0) \rangle.
\end{equation}
We can relate  $C_{{\cal V},-,+} (\pp)$  to the OPE coefficient
$C^{\cal V}_{-,+} (-\pp)$ using (\ref{sigma-sigma-OPE-2}) with $x=1$:
\begin{eqnarray}
\label{reluplow}
C_{{\cal V},-,+} (\pp)
&=&  \int \frac{d^2 p'}{(2 \pi)^2} \,
C^{\cal V}_{-,+} (\pp') \,
\langle\,  {\cal V}_\pp (\infty) \, {\cal U}_{\pp'} (0) \,\rangle
\nonumber \\
&=& \frac{1}{N} \sum_{i=0}^{N-1}
\int \frac{d^2 p'}{(2 \pi)^2} \, C^{\cal V}_{-,+} (\pp') \,
\langle \, {\cal U}_{\alpha^{i} \pp} (\infty) \,
{\cal U}_{\pp'} (0) \,\rangle\,,\nonumber \\
&=& \frac{1}{N} \sum_{i=0}^{N-1}
C^{\cal V}_{-,+} (- \alpha^{i} \,\pp)\, V_{D-2}
=  C^{\cal V}_{-,+} (-\pp)\,   V_{D-2}\, .
\end{eqnarray}
The three-point function  $C_{{\cal V},-,+}
(\pp)$ is effectively the one-point function of an operator
of the form $\exp (i \pp \XX + i \bar{\pp} \bar{\XX})$,
with twisted boundary conditions.
It is given by an exponential of the propagator
$- \bar{\pp} \pp \, \langle \bar{\XX} \XX \rangle$
evaluated with twisted boundary conditions,
so it has a Gaussian dependence on the momentum.
Since the overall normalization of $C_{{\cal V},-,+} (\pp)$
is fixed by (\ref{normatzeromom}) and (\ref{reluplow}),
we can write
\begin{equation}
\label{theansatz}
C_{{\cal V},-,+} (\pp)
=  V_{D-2} \, {\delta'}^{-\alpha' \bar{\pp} \pp}, \quad
C^{\cal V}_{-,+} (\pp) = {\delta'}^{-\alpha' \bar{\pp} \pp},
\end{equation}
where $\delta'$ is a constant to be determined.

\medskip
For the two-dimensional orbifold of infinite volume,
the correlator of four twist operators
is given in~\cite{Dixon:1986qv}:
\begin{equation}
Z (x, \bar{x})
= \langle \, \sigma_{-} (\infty) \, \sigma_{+} (1) \,
\sigma_{-} (x, \bar{x}) \, \sigma_{+} (0) \rangle
= \frac{{\cal N}
\, |x(1-x)|^{-\frac{2k}{N} \left( 1 - \frac{k}{N} \right)}}
{F(x) F(1-\bar{x}) + F(1-x) F(\bar{x})} \,,
\end{equation}
where $F(x) \equiv F ( \frac{k}{N}, 1-\frac{k}{N}; 1; x)$
is a hypergeometric function.
To determine the normalization constant ${\cal N}$ we examine
the behavior of $Z (x, \bar{x})$ when $x \sim 0$.
On the one hand, using the OPE (\ref{sigma-sigma-OPE}) and
(\ref{theansatz}) we find
\begin{eqnarray}
Z (x, \bar{x})
&\sim& \int_\Gamma \frac{d^2 p }{(2 \pi)^2} \,
\, N\, C^{\cal V}_{-,+} (\pp) \, |x|^{2 \alpha' \bar{\pp} \pp
- \frac{2k}{N} \left( 1 - \frac{k}{N} \right) } \,
\langle \, \sigma_{-} (\infty) \, \sigma_{+} (1) \,
{\cal V}_\pp (0) \rangle
\nonumber \\
&=& N\int_\Gamma \frac{d^2 p }{(2 \pi)^2} \,
C^{\cal V}_{-,+} (\pp) \, C_{{\cal V},-,+} (\pp) \,
|x|^{2 \alpha' \bar{\pp} \pp
- \frac{2k}{N} \left( 1 - \frac{k}{N} \right) }
\nonumber \\
&=&  V_{D-2}\, |x|^{- \frac{2k}{N} \left( 1 - \frac{k}{N} \right)}
\int \frac{d^2 p }{(2 \pi)^2} \,
\left| \frac{x}{\delta'} \right|^{2 \alpha' \bar{\pp} \pp}
\nonumber\\
&=& \frac{1}{2 \pi \alpha'}\,
\left( -\ln \left| \frac{x}{\delta'} \right| \right)^{-1}
|x|^{- \frac{2k}{N} \left( 1 - \frac{k}{N} \right)} \, V_{D-2}\,.
\end{eqnarray}
On the other hand, for $x\sim 0$,  $F(x)\sim 1$ and
$F(1-x)\sim\frac{1}{\pi} \sin \left( \frac{\pi k}{N} \right)
\left( -\ln \frac{x}{\delta} \right)$. Here $\delta$ is a
function of $k/n$:
$\ln \delta \left( k/N\right)
= 2 \, \psi (1) - \psi \left( k/N\right)
- \psi ( 1- (k/N))$,
where $\psi (x) = \Gamma'(x)/\Gamma(x)$.
In particular, $\delta(1/2) = 2^4$ and $
\delta (1/3) = 3^3$.
We then find
\begin{equation}
Z (x, \bar{x}) \sim {\cal N}
\left[ \frac{2}{\pi} \sin \left( \frac{\pi k}{N} \right)
\left( -\ln \left| \frac{x}{\delta} \right| \right) \right]^{-1}
|x|^{- \frac{2k}{N} \left( 1 - \frac{k}{N} \right)}.
\end{equation}
By comparing the two expressions of $Z (x, \bar{x})$
when $x \sim 0$, we deduce that
\begin{equation}
\delta' = \delta\,,
\end{equation}
which fixes the three-point functions in (\ref{theansatz}).
The value of   ${\cal
N}$ is also determined, and  the result for
$Z (x,
\bar{x})$ is
\begin{equation}
\label{finreszxxbar}
Z (x, \bar{x}) = \frac{1}{\pi^2 \alpha'}\,
\sin \left( \frac{\pi k}{N} \right)
\frac{|x(1-x)|^{-\frac{2k}{N} \left( 1 - \frac{k}{N} \right)}}
{F(x) F(1-\bar{x}) + F(1-x) F(\bar{x})}\,\,
V_{D-2}\, .
\end{equation}

\medskip
We now use the $x \to \infty$ behavior of $Z (x, \bar{x})$
to calculate the three-twist correlators
\begin{equation}
C_{--,+,+} = \langle \sigma_{--} (\infty) \,
\sigma_{+} (1) \, \sigma_{+} (0) \rangle, \quad
C_{++,-,-} = \langle \sigma_{++} (\infty) \,
\sigma_{-} (1) \, \sigma_{-} (0) \rangle \,.
\end{equation}
The coefficients $C^{++}_{+,+}$ and $C^{--}_{-,-}$ are
defined by the leading terms in the OPE's,
\begin{equation}
\sigma_{+} (x, \bar{x}) \, \sigma_{+} (0) \sim C^{++}_{+,+} \,
|x|^{-\frac{2 k^2}{N^2}} \, \sigma_{++} (0), \quad
\sigma_{-} (x, \bar{x}) \, \sigma_{-} (0) \sim C^{--}_{-,-} \,
|x|^{-\frac{2 k^2}{N^2}} \, \sigma_{--} (0)\,,
\end{equation}
and they and the three-twist correlators
obey the following relations:
\begin{equation}
C_{--,+,+} = C_{++,-,-} = C^{++}_{+,+} V_{D-2} = C^{--}_{-,-} \,
V_{D-2}\,.
\end{equation}
On the one hand, for $x \to \infty$, we find
\begin{equation}
Z (x, \bar{x}) \,\sim\, C^{--}_{-,-} \,
|x|^{-\frac{2k}{N} \left(1-\frac{2k}{N}\right)} \,
\langle \, \sigma_{--} (\infty) \,
\sigma_{+} (1) \, \sigma_{+} (0) \rangle
=  C^{--}_{-,-} \, C_{--,+,+} \,
|x|^{-\frac{2k}{N} \left(1-\frac{2k}{N}\right)}.
\end{equation}
On the other hand, for $x \to \infty$
the asymptotic forms of $F(x)$ and $F(1-x)$
\begin{equation}
F(x) \sim \alpha^{\frac{k}{2}} \,
\frac{\Gamma \left( 1-\frac{2k}{N} \right)}
{\Gamma^2 \left( 1-\frac{k}{N} \right)} \, x^{-\frac{k}{N}}, \quad
F(1-x) \sim \frac{\Gamma \left( 1-\frac{2k}{N} \right)}
{\Gamma^2 \left( 1-\frac{k}{N} \right)} \, x^{-\frac{k}{N}},\quad
k/N < 1/2\,,
\end{equation}
immediately give the $x \to \infty$ behavior of  $Z (x, \bar{x})$:
\begin{equation}
Z (x, \bar{x}) \sim
\frac{1}{2 \pi^2 \alpha'} \tan \left( \frac{\pi k}{N} \right)
\frac{\Gamma^4 \left( 1-\frac{k}{N} \right)}
{\Gamma^2 \left( 1-\frac{2k}{N} \right)} \,
|x|^{-\frac{2k}{N} \left(1-\frac{2k}{N}\right)}\, V_{D-2}, \quad
k/N<1/2\,.
\end{equation}
Comparing the two asymptotic formulae for $Z (x, \bar{x})$ given
above, we find that
\begin{equation}
\label{answerthreepoint}
C_{--,+,+} = C_{++,-,-}
= \sqrt{ \frac{1}{2 \pi^2 \alpha'} \tan \frac{\pi k}{N} } \,\,
\frac{\Gamma^2 \left( 1-\frac{k}{N} \right)}
{\Gamma \left( 1-\frac{2k}{N} \right)}\, V_{D-2}\, , \quad
k/N < 1/2\,.
\end{equation}
This result can be compared with equation (4.47) of
\cite{Dixon:1986qv},
which holds for
an orbifold with finite two-dimensional volume
$V_\Lambda$.
The answer in (\ref{answerthreepoint}) is obtained
by discarding the factor
which involves the sum over classical solutions
that exist for finite orbifolds
and by replacing $V_\Lambda$
by $1/(2\pi^2)$.  Additionally, (\ref{answerthreepoint})
contains the $\alpha'$ dependence
necessary to get tachyon potentials with the correct units
and the spacetime volume factor $V_{D-2}$.
When the operators in any correlator
that has this volume factor are dressed
with $D-2$ dimensional momenta $q$,
the factor must be replaced
by $(2\pi)^{D-2} \delta^{(D-2)} (\sum q)$.

\sectiono{The $\mathbb{C}/\mathbb{Z}_2$ tachyon potential}
\label{secz2pot}

Having computed the necessary CFT correlators,
we can formulate closed string field theory on
orbifold  backgrounds.
In this section we study the string field theory of
the simplest one, the
$\mathbb{C}/\mathbb{Z}_2$ orbifold.
We first compute the closed string field theory action,
and then we carry out the level-truncation analysis.

\subsection{Computing the action}

The $\mathbb{C}/\mathbb{Z}_2$ orbifold consists of two sectors:
an untwisted sector corresponding to $k=0$ and
a twisted sector with $k=1$.
The structure of the untwisted sector is universal for any
$\mathbb{C}/\mathbb{Z}_N$ orbifold.
The vertex operator for the bulk
tachyon is given by (\ref{bulk-tachyon-vertex-operator})
and we write the corresponding string field as
\begin{equation}
|U\rangle = \int \frac{d^{D-2} q}{(2 \pi)^{D-2}}
\int \frac{d^2 p}{(2 \pi)^2} \,
u(q,p) \, c_1 \bar{c}_1 |q,p\rangle\,.
\end{equation}
Since the correlators
of $c\bar ce^{iq\cdot X} \mathcal{U}_\pp$ are identical to those
of $c\bar ce^{iq\cdot X} e^{i p_1 X^1 + i p_2 X^2}$
in the flat-space theory,
the tachyon string field action takes the same form
as~(\ref{utachefffac}):
\begin{eqnarray}
&& -\frac{1}{2} \int \frac{d^{D-2} q}{(2 \pi)^{D-2}}
\int\frac{d^2 p}{(2 \pi)^2} \,
u(-q,-p) \Bigl( q^2 + p^2 -\frac{4}{\alpha'} \Bigr)
u(q,p)\nonumber\\
&& -\frac{1}{3!} \frac{4 \kappa}{\alpha'}
\int \prod_{i=1}^{3} \Bigl[ \frac{d^{D-2} q_i}{(2 \pi)^{D-2}}
\frac{d^2 p_i}{(2 \pi)^2} \,
\RR^{2 -\frac{1}{2} \alpha' q_i^2 -\frac{1}{2} \alpha' p_i^2}
u(q_i,p_i) \Bigr]
\nonumber \\ && \qquad \qquad \times
(2\pi)^D \delta^{(D-2)} ( \, q_1 + q_2 + q_3)
\delta^{(2)} (p_1 + p_2 + p_3)\,.
\end{eqnarray}
The only $N$ dependence that remains in the above terms is
due to the condition $u(q, \pp) = u(q, e^{2\pi i/N} \pp)$.
It is convenient to
make a Fourier transformation on each of the flat $D-2$ directions:
\begin{eqnarray}
&& -\frac{1}{2} \int d^{D-2} x \int \frac{d^2 p}{(2 \pi)^2}
\Bigl[ \eta^{\mu \nu} \, \partial_\mu u(x,-p) \,
\partial_\nu u(x,p)
{}+ u(x,-p) \Bigl( p^2 -\frac{4}{\alpha'} \Bigr)
u(x,p) \Bigr] \nonumber\\
&& -\frac{1}{3!} \frac{4 \kappa}{\alpha'}
\int d^{D-2} x \int \prod_{i=1}^{3} \Bigl[
\frac{d^2 p_i}{(2 \pi)^2} \,
\RR^{2 +\frac{1}{2} \alpha' \partial^2 -\frac{1}{2} \alpha' p_i^2}
u(x,p_i) \Bigr]
(2\pi)^2 \delta^{(2)} (p_1 + p_2 + p_3) \,.
\end{eqnarray}
For configurations that are translationally invariant
along the $(D-2)$-dimensional spacetime we can define
a potential $\mathbb{V}$ that is equal
to minus the $(D-2)$-dimensional Lagrangian density.
The untwisted tachyon contribution to the potential is therefore
\begin{eqnarray}
\mathbb{V}_{u^2} + \mathbb{V}_{u^3}
&=& -\frac{1}{2} \int \frac{d^2 p}{(2 \pi)^2} \,
u(-p) \Bigl( \frac{4}{\alpha'} - p^2 \Bigr) u(p)
\nonumber \\
&& {}+ {1\over 3 !} {4\kappa\over \alpha'}
\int \prod_{i=1}^{3} \Bigl[
\frac{d^2 p_i}{(2 \pi)^2}\,\RR^{2 -\frac{1}{2} \alpha' p_i^2} \,
u(p_i) \Bigr] (2\pi)^2 \delta^{(2)} (p_1+ p_2 + p_3) \,.
\label{E_u^2+E_u^3}
\end{eqnarray}

Let us next consider the twisted sector.
The twist operator, denoted by $\sigma$, has
conformal dimensions  $h_{\sigma}=\bar h_\sigma = 1/8$.
The associated  twisted tachyon field $t(x)$ has
mass-squared equal to  ${2/ \alpha'}$
times the sum of the holomorphic and antiholomorphic dimensions of
$c_1 \bar c_1 | \sigma \rangle$: $m^2 = {2\over \alpha'}
\left( -1 + {1\over 8} -1 + {1\over 8} \right)
= {4\over \alpha'} \left( -{7\over 8}\right)$.
The level of this field is obtained using (\ref{cstaprimsdnlevvel}):
\begin{equation}
\label{levelt}
\ell(t) ={1\over 4}\,.
\end{equation}
The next excited states in the twisted sector
contain two matter oscillators
each of which contributes level $1/2$,
so the level of the corresponding tachyon states is $5/4$.
Our present analysis will  not include these states,
whose quadratic terms appear at level $10/4$, but only the bulk
tachyon and the twisted tachyon $t$.
The string field associated with $t$ is written as
\begin{equation}
|T\rangle
= \int {d^{D-2}q\over (2\pi)^{D-2}} \,
t(q) \, c_1 \bar c_1 |\sigma, q\rangle \,, \quad
t(q) = \int d^{D-2} x  \, t (x) \, e^{-iq\cdot x}\,.
\end{equation}
Since both BPZ and Hermitian conjugations
take $| \sigma \rangle$
to the same bra $\langle \sigma |$,
the string field is real if $t(x)$ is real.
The inner product is given by
\begin{eqnarray}
\langle\sigma, -q' | \, c_{-1} \bar c_{-1} \,\, c_0^- c_0^+ \,\,
c_1 \bar c_1 | \sigma , q\rangle &=& (2\pi)^{D-2} \, \delta^{(D-2)}( 
q+ q') \,,\\
\langle\sigma, -q' | \, c \bar c (\infty) \,\, c\bar c (1) \,\,
c\bar c (0) | \sigma , q\rangle &=& 2\, (2\pi)^{D-2} \, 
\delta^{(D-2)}( q+ q')\,,
\end{eqnarray}
as expected for a twisted field that lives
in $D-2$ spacetime dimensions.
The kinetic term for the twisted tachyons is found to be
\begin{equation}
- \frac{1}{2} \int  {d^{D-2}q\over (2\pi)^{D-2}} \,
t (-q) \Bigl( q^2 - {4\over \alpha'}{7\over 8}\Bigr)  t (q) \,.
\end{equation}
The associated contribution
to the potential is
\begin{equation}
\mathbb{V}_{t^2}
= -\frac{1}{2} \, \frac{4}{\alpha'} \, \frac{7}{8} \, t^2 \,.
\end{equation}

There is no $t^3$ interaction
because of the $\mathbb{Z}_2$ symmetry of the orbifold CFT.
There is, however, a cubic coupling
that involves a bulk tachyon and two twisted tachyons.
Using the formula (\ref{evalS3})
this interaction is given by the following term in the action:
\begin{eqnarray}
&&-{1\over 3!} \,{2\kappa\over \alpha'} \cdot 3\cdot \int
\prod_{i=1}^3 \Bigl[ {d^{D-2} q_i\over (2\pi)^{D-2}}\,
\RR^{2-{1\over 2} \alpha' q_i^2}\Bigr] \int{d^2 p\over
(2\pi)^2}  \RR^{-{1\over 2} \alpha' p^2} u(q_1, p)\,
\RR^{-{1\over 4}} t(q_2) \,\RR^{-{1\over 4}} t(q_3) \nonumber \\
&& \qquad {}\times 2\,
\delta^{-{1\over 4} \alpha' p^2} (2\pi)^{D-2}
\delta^{(D-2)}(q_1 + q_2 + q_3)\,,
\end{eqnarray}
where the factor on the second line is the three-point correlator
$C_{\mathcal{V}, -,+}(\pp )$
in (\ref{theansatz}), extended to the case where the operators are
dressed with $(D-2)$-dimensional momenta and the ghosts
are included.
After Fourier transformation and simplification,
the above term becomes
\begin{equation}
-\frac{2 \kappa}{\alpha'} \,
\RR^{\frac{11}{2}}
\int d^{D-2} x \int \frac{d^2 p}{(2 \pi)^2}
\left( \RR^2 \delta \right)^{-\frac{1}{4} \alpha' p^2}
\bigl[ \RR^{\frac{\alpha' \partial^2}{2}}
u(x,p) \bigr]
\bigl[ \RR^{\frac{\alpha' \partial^2}{2}}
t (x) \bigr]^2 \,.
\label{ut^2}
\end{equation}
The contribution
to the potential from this coupling is
\begin{equation}
\label{cubutsquared}
\mathbb{V}_{u t^2} = \frac{2 \kappa}{\alpha'} \,
\RR^{\frac{11}{2}} \, t^2
\int \frac{d^2 p}{(2 \pi)^2}
\left( \RR^2 \delta \right)^{-\frac{1}{4} \alpha' p^2}
u(p)  \,, \quad
\delta = 2^4 ~~ \text{for} ~~ \mathbb{C}/\mathbb{Z}_2 \,.
\end{equation}
Collecting all the contributions,
the potential is given by
\begin{eqnarray}
\mathbb{V}_2 &=& - \frac{1}{2} \,
\frac{4}{\alpha'} \, \frac{7}{8} \, t^2-\frac{1}{2} \int \frac{d^2
p}{(2 \pi)^2} \,
u(-p) \left( \frac{4}{\alpha'} - p^2 \right) u(p)  {}
\nonumber \\[0.5ex] &&
+ \frac{2 \kappa}{\alpha'} \,
\RR^{\frac{11}{2}}\, t^2\,
\int \frac{d^2 p}{(2 \pi)^2}
\left( \RR^2 \delta \right)^{-\frac{1}{4} \alpha' p^2} u(p) \,
\nonumber \\
&& {}+ {1\over 3 !} {4\kappa\over \alpha'}
\int \prod_{i=1}^{3} \left[
\frac{d^2 p_i}{(2 \pi)^2}\,\RR^{2 -\frac{1}{2} \alpha' p_i^2} \,
u(p_i) \right] (2\pi)^2 \delta^{(2)} (p_1+ p_2 + p_3) \,.
\end{eqnarray}
In order to eliminate $\alpha'$ and $\kappa$
we define dimensionless momentum variables $(\xi_1, \xi_2)$
and conjugate coordinates $(r_1, r_2)$ by
\begin{equation}
\label{scaledxip}
p_a = \frac{2}{\sqrt{\alpha'}} \, \,\xi_a \,, \quad
x_a = {\sqrt{\alpha'}\over 2} \, r_a\, , \quad a=1,2\,,
\end{equation}
where  $x_a$ is the coordinate conjugate to $p_a$.
Additionally, we define
\begin{equation}
\label{rescalingup}
u(\xi) = \frac{4 \kappa}{\alpha'} \, u(p)
\end{equation}
and rescale $t \to \frac{\sqrt{\alpha'}}{\kappa} \, t$.
The resulting expression for the
ratio $f_2$ defined by (\ref{f_N-definition}) is
\begin{eqnarray}
\label{f2momspace}
f_2 = \frac{\kappa^2 \, \mathbb{V}_2}{\pi}
&=& - \frac{7}{4 \pi} \, t^2
- \frac{1}{2 \pi} \int \frac{d^2 \xi}{(2 \pi)^2} \,
u(-\xi) \left( 1 - \xi^2 \right) u(\xi)
\nonumber \\
&& {} + \frac{2}{\pi} \, \RR^{\frac{11}{2}} \, t^2 \,
\int \frac{d^2 \xi}{(2 \pi)^2}
\left( \RR^2 \delta \right)^{-\xi^2} u(\xi)
\nonumber \\
&& {}+ \frac{1}{6 \pi}
\int \prod_{i=1}^{3} \Bigl[
\frac{d^2 \xi_i}{(2 \pi)^2}\,\RR^{2 (1 - \xi_i^2)} \,
u(\xi_i) \Bigr] (2\pi)^2 \delta^{(2)} (\xi_1+ \xi_2 + \xi_3) \,.
\end{eqnarray}

\begin{figure}[thb]
\centerline{\epsfxsize=3in\epsfbox{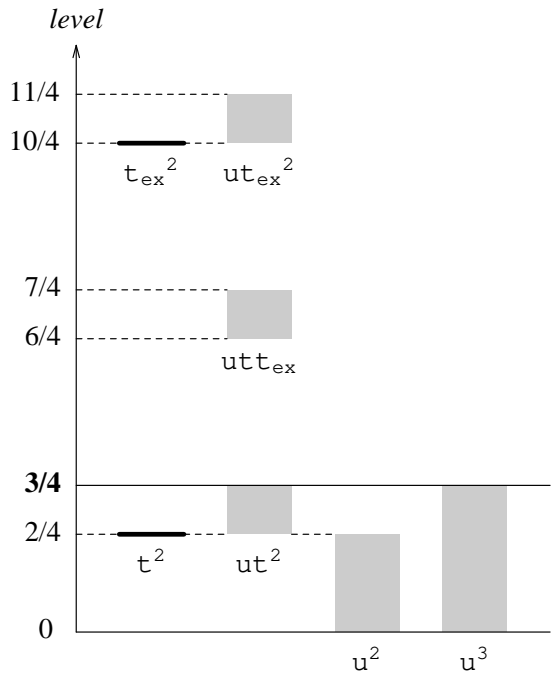}}
\caption{\small
Level expansion table for the $\mathbb{C}/\mathbb{Z}_2$ orbifold.
Since the bulk tachyon includes all levels from zero to $1/4$,
interactions that involve the bulk tachyon are shown as bands.
Terms involving excited twisted tachyons $t_{\rm{ex}}$ are shown,
but they are not included in our analysis.}
\label{levexptabz2}
\end{figure}

We will be analyzing $f_2$
in the level expansion.  The level of the
untwisted tachyon $u(p)$ is given by (\ref{cstachyonlevvel}):
\begin{equation}
\ell \left( u(p) \right) = {1\over 2} \alpha'p^2 = 2 \xi^2\,.
\end{equation}
Working up to some level means including only a limited set
of momentum modes of the field $u(p)$. This is exactly the
way that open string tachyon modes were treated in the case of
spatially dependent lump solutions~\cite{Moeller:2000jy}.
As we discussed earlier, the level of the twisted field $t$
is equal to $1/4$, so it is reasonable to use all bulk tachyon modes
that have level less than or equal to $1/4$.
We therefore demand
\begin{equation}
\label{levconzw2}
\ell (u(\xi)) = 2\xi^2 \leq {1 \over 4} \, \quad \to
\quad |\xi| \leq \xi_* = \sqrt{\frac{1}{8}} \simeq 0.354 \,.
\end{equation}
As a result, in the level expansion analysis of $f_2$, the integrals
must be cutoff at
$\xi=\xi_*$.
The levels of the various fields and interactions are represented
in Fig.~\ref{levexptabz2}.  With maximum level $1/4$ for
$u(p)$, the cubic self-interaction $u^3$ has level less than or equal
to $3/4$. For illustration
purposes only, this figure includes couplings of the excited
twisted tachyons $t_{\rm{ex}}$ that appear
at levels higher than the cutoff level~$3/4$.

\medskip
To study $f_2$ using coordinate space
we define $u(r)$ by standard two-dimensional Fourier transformation:
\begin{equation}
\label{dflkj}
u(\xi) = \int d^2r  \, u(r)\, e^{-i \xi\cdot r} \,.
\end{equation}
The relation between $u(p)$ and $u(x)$
in (\ref{bulk-tachyon-string-field})
together with (\ref{scaledxip}), (\ref{rescalingup}), and
(\ref{dflkj}) implies that
$u(x) = u(r)/\kappa$.
We use (\ref{dflkj}) and the rescaled $t$
to rewrite $f_2$  in coordinate space. Since
we are only interested in rotationally invariant configurations of
the field $u$,
we find
\begin{equation}
\label{coordspacefunctional}
f_2 = - \frac{7}{4 \pi} \, t^2+\int_0^\infty\hskip-4pt r dr
\Bigl( - u(r) \left( \partial^2 + 1 \right) u(r)
+\frac{\RR^{\frac{11}{2}} t^2}{\pi \ln (\RR^2 \delta)} \,
e^{-\frac{r^2}{4 \ln (\RR^2 \delta)}} \, u(r)
+ \frac{1}{3} \left[ \RR^{2 (\partial^2+1)} u(r) \right]^3
\Bigr)\,,
\end{equation}
where  $r= \sqrt{r_1^2 + r_2^2}$, and $\partial^2 =
\frac{\partial^2}{\partial r^2}
+ \frac{1}{r} \frac{\partial}{\partial r} $ is the two-dimensional
Laplacian that acts on rotationally invariant functions. Equation
(\ref{coordspacefunctional}) will be the starting point of our
coordinate space analysis.

\subsection{A preliminary analysis}\label{prelimanalesvlkd}

We have presented an expression (\ref{f2momspace})
for $f_2$ that incorporates a restricted set of interactions.
We must now find the critical points of $f_2$
in the configuration space of $t$ and  $u(\xi)$.
If the value of $f_2$ at a critical point is
near minus one, this can be taken as evidence for the conjecture
that gives the height of the potential in terms of the deficit
angle of the cone.

The equations that define  critical points are made very
nontrivial by the cubic term of the bulk tachyon $u$.
Let us first study $f_2$
in the approximation where the $u^3$ term is ignored.
We will take $f_2$ to be given by
\begin{eqnarray}
\label{f2momtrsp}
f_2= - \frac{7}{4\pi} \, t^2 -\frac{1}{2\pi} \int \frac{d^2 \xi}
{(2 \pi)^2} \, u(-\xi) \left( 1 - \xi^2 \right) u(\xi)
{}+\, \frac{2}{\pi} \,
\RR^{\frac{11}{2}}
\, t^2\,  \int \frac{d^2 \xi}{(2 \pi)^2}
\left( \RR^2 \delta \right)^{-\xi^2} u(\xi) \,,
\end{eqnarray}
and we will determine the critical points of this potential
analytically.

It is in fact instructive to obtain an effective potential for~$t$
by solving for $u(\xi)$ in terms of $t$ and substituting back into
$f_2$. Varying $f_2$ with respect to
$u(\xi)$ one readily finds that
\begin{equation}
u(\xi) = 2 \RR^{\frac{11}{2}} \,
\frac{(\RR^2 \delta)^{-\xi^2}}{1-\xi^2} \, t^2\,.
\end{equation}
Substituting back into (\ref{f2momtrsp}), the effective potential
for $t$ acquires a positive $t^4$ term:
\begin{equation}
\label{thepoteff}
f_2 = - \frac{7}{4\pi} \, t^2 + {2 \over \pi} \,
\eta (\RR,\delta, \xi_*)\, \RR^{11} \, t^4 \,,
\end{equation}
where $\eta (\RR,\delta, \xi_*)$ is a constant defined
by the following integral:\footnote{
Actually, $ \eta (\RR,\delta, \xi_*) = (4 \pi \, \RR^4
\delta^2)^{-1} [ {\rm Ei} ( 2 \ln ( \RR^2 \delta) )
-{\rm Ei} ( 2 \, (1-\xi_*^2) \ln ( \RR^2 \delta))]\,,$
where the exponential integral function
${\rm Ei} \,
(x) = -\int_{-x}^{\infty} dt \, \frac{e^{-t}}{t}$
is defined by taking the principal value at $t=0$.}
\begin{equation}
\label{defenetansl}
\eta (\RR,\delta, \xi_*) = \int_0^{\xi_*} d \xi \, \frac{\xi}{2 \pi}
\frac{(\RR^2 \delta)^{-2 \, \xi^2}}{1-\xi^2}
\simeq 0.00717348 \,,
\end{equation}
evaluated with $\delta = 2^4$ and $\xi_* = \sqrt{1/8} \,$.
While $\RR$ is a constant, we include it as an argument of $\eta$
to discuss the $\RR$  dependence of the results explicitly.
The potential (\ref{thepoteff}) has two critical points
with the same value of $f_2$:
\begin{equation}
\label{f2valsft}
f_2 = -\frac{49}{128 \pi} \,
{\RR^{-11} \over \eta (\RR,\delta, \xi_*)} \simeq -0.955577 \,.
\end{equation}
This answer is remarkably close to the expected value!
We will see in the following subsections
that the inclusion of the cubic bulk
tachyon term modifies the result, but not drastically.
The cubic term must be included, of course, in any systematic
analysis that aims to be fully quantitative.

Had we not used the closed string field theory action,
but rather the naive extrapolation from on-shell CFT computations,
$f_2$ would be given by the expression in  (\ref{f2valsft})
evaluated for $\RR = 1$. It turns out that $\eta$ is a relatively
slowly varying function of $\RR$: for $\RR=1$, $\eta
\simeq 0.00760665$, which is about 6\% larger than the value
given in (\ref{defenetansl}).
Therefore the change in $f_2$ is mostly due to the factor
$\RR^{-11} \simeq 1/17.78$.
Indeed, for $\RR=1$, one finds $f_2\simeq -16.0193$,
which is far away from the expected result.
This demonstrates that quantitative study requires
the use of the off-shell CSFT action.

This value of $f_2$ does not change much when we increase
the momentum cutoff $\xi_*$. Increasing the cutoff is equivalent
to increasing the level $\ell$ of the allowed bulk tachyon modes.
It is clear from Fig.~\ref{levexptabz2} that it is not really
consistent to allow arbitrarily high-level tachyon modes
without including other fields and interactions.
We do it, however, to get a sense of the effects
of higher bulk tachyon modes.  While $\xi_*$
is less than one and increasing, $\eta$ increases,
the quartic term becomes larger,
and the critical point becomes shallower.
If $\xi_*>1$, which corresponds to level $\ell >2$,
the integral for $\eta$ requires a prescription
to deal with the pole at $\xi=1$.
If we take the principal value at $\xi=1$,
the value of $f_2$ remains essentially constant
for all  $\xi_* > 1$ and $f_2 \simeq -0.456693$
in the limit $\xi_* \to \infty$.
The singularity at $\xi=1$ arises because
the bulk tachyon becomes marginal at this momentum.
The coordinate space solution develops a long-ranged
Bessel function tail,
and a detailed study shows that
the principal value prescription gives
a solution with minimal tail.

\subsection{Level expansion analysis}

We now investigate the critical points of the functional $f_2$
including the cubic term of the bulk tachyon.
It is no longer possible
to solve the equations of motion analytically,
so we discretize the spectrum of the bulk tachyon
and solve the equations numerically.
We explore two different methods
for the discretization,
in both of which the level is constrained as in (\ref{levconzw2}).
In the first one we use coordinate space:  we replace the
two-dimensional space of the cone by a round disk
and impose the condition that the bulk tachyon vanishes
at the boundary.   In the second one we do a straightforward
discretization of the momentum space representation of
$f_2$.  By doing so, the two-dimensional space of the cone
has been replaced by a square box with periodic boundary
conditions.
Since we are looking for a solution where the bulk tachyon
is localized near the origin,  we expect
both methods to give the same result. Indeed, this is what
we~find.

\subsubsection{Coordinate space method}\label{dlfkjgoin}

In this approach we study $f_2$ (see (\ref{coordspacefunctional}))
by imposing the condition $u(r=L)=0$, where $L$ is a length
to be fixed by the desired accuracy of the analysis.
We will expand $u(r)$
using Bessel functions $J_0 ( \xi \, r)$ that vanish for $r=L$.
We can view $J_0(\xi\, r)$ as a superposition of plane waves,
\begin{equation}
J_0 ( \xi \, r) = \frac{1}{2 \pi} \int_0^{2 \pi} d \theta \,
e^{i \, \xi \, r \cos \theta} \,,
\end{equation}
each of which  carries momentum of magnitude $\xi$.  Indeed,
the two-dimensional Laplacian  gives
$\partial^2J_0(\xi\, r) = -\xi^2 J_0(\xi\, r)$.
We therefore expand the bulk tachyon $u(r)$ in terms of
functions $J_0(\xi\, r)$  with $\xi \le \xi_*$.
We can select $n$ such functions
\begin{equation}
\{ J_0(\xi_1\, r)\,, ~J_0(\xi_2\, r)\,, ~  \ldots \, ~
J_0(\xi_n\, r) \} \, \quad\text{with} \quad
\xi_1 < \xi_2 < \ldots < \xi_n = \xi_*,
\end{equation}
by imposing the $n$ conditions $\xi_i L = \alpha_i$, where
$\alpha_i$ denotes the $i$-th zero of $J_0(x)$:  $J_0 (\alpha_i)=0$.
We use the condition with $i=n$ to fix the size $L$ of the disk and
the remaining conditions to fix the values of~$\xi_i$:
\begin{equation}
L = {\alpha_n\over \xi_*} \,, \quad ~ \xi_i = {\alpha_i\over L } =
{\alpha_i\over \alpha_n} \,\xi_* \,.
\end{equation}
The bulk tachyon $u(r)$ is thus expanded as
\begin{equation}
u(r) = \sum_{i=1}^n u_i \, \psi_i (r) \,, \quad
\text{with} \quad \psi_i (r)
= \frac{\xi_*}{\sqrt{\pi} \, \alpha_n J_1 (\alpha_i)}
J_0 \left(\alpha_i \, {r\over L} \right) \,,
\end{equation}
which are functions  normalized to satisfy
\begin{equation}
2 \, \pi \int_0^L dr \, r \,
\psi_{i} (r) \, \psi_{j} (r) = \delta_{ij} \,.
\end{equation}
At this $n$-th order approximation,
$f_2$ is  a function of the $n$ modes
$u_i$ of the bulk tachyon and the twisted tachyon  $t$.
The continuum spectrum is recovered in the limit $n \to \infty$.

Let us now turn to the explicit computations.
Many of the general features of the minimization of $f_2$
can be gleaned from the simplest case $n=1$,
which we explore first.
Although the resulting values of $f_2$ differ
substantially from those obtained for large $n$,
the $n=1$ case can be studied analytically.
We then turn to the large $n$ numerical analysis.

\medskip
\noindent
$\underbar{Lowest-order approximation}$.
A simple computation shows that for $n=1$, $f_2$ is given by
\begin{equation}
f_2 = - \frac{1}{2} \, m_u^2 \, u_1^2
- \frac{1}{2} \, m_t^2 \, t^2
+ g_{u^3} \, u_1^3 + g_{u t^2} \, u_1 \, t^2 \,,
\end{equation}
where
\begin{eqnarray}
&& m_u^2 = \frac{1-\xi_*^2}{\pi} \simeq 0.278521 \,,
\quad m_t^2 = \frac{7}{2 \, \pi} \simeq 1.11408 \,,
\nonumber \\ &&
g_{u^3} = \frac{\xi_* \, \RR^{6 \, (1-\xi_*^2)}}
{3 \, \pi^{\frac{3}{2}} \, \alpha_1 \, J_1 (\alpha_1)^3}
\int_0^1 dy \, y \, J_0 (\alpha_1 \, y)^3
\simeq 0.0242103 \,,
\nonumber \\ &&
g_{u t^2} = \frac{\alpha_1 \, \RR^{\frac{11}{2}}}
{\pi^{\frac{3}{2}} \, \xi_* \,
J_1 (\alpha_1) \, \ln (\RR^2 \delta)}
\int_0^1 dy \, y \, e^{-\frac{\alpha_1^2}
{4 \, \xi_*^2 \, \ln (\RR^2 \delta)} \, y^2}
J_0 (\alpha_1 \, y)
\simeq 0.285768 \,.
\end{eqnarray}
Although $f_2$ contains four parameters,
we can eliminate two of them by rescaling  $u_1$ and  $t$,
and we can eliminate a third one  by rescaling $f_2$ itself:
\begin{equation}
\label{fullpotlev}
\widetilde{f}_2 = -\frac{1}{2} \, \tilde{u}_1^2
-\frac{1}{2} \, \tilde{t}^{\, 2} + \frac{\GG}{3} \, \tilde{u}_1^3
+ \tilde{u}_1 \, \tilde{t}^{\, 2} \,,
\end{equation}
where
\begin{equation}
\widetilde{f}_2
= \frac{g_{u t^2}^2}{m_u^2 \, m_t^4} \, f_2 \,, \quad
\tilde{u}_1 = \frac{g_{u t^2}}{m_t^2} \, u_1 \,, \quad
\tilde{t} = \frac{g_{u t^2}}{m_u \, m_t} \, t \,,
\quad
\GG = \frac{3 \, g_{u^3} \, m_t^2}{g_{u t^2} \, m_u^2} \,.
\end{equation}
This form is useful because the vacuum structure
is governed only by $\GG$, which in our case is given by
\begin{equation}
\label{rvalue}
\GG \simeq 1.01664 \,.
\end{equation}
Solving the equations $\partial \widetilde{f}_2/\partial
\tilde{u}_1 =\partial
\widetilde{f}_2/\partial \tilde{t} = 0$,
we find three nontrivial critical points:
\begin{eqnarray}
&& \widetilde{f}_2 = - \frac{3-\GG}{24} \quad \text{for} \quad
( \tilde{u}_1, \, \tilde{t} \, )
= \left( \frac{1}{2}, \, \pm \frac{\sqrt{2-\GG}}{2} \right), \,
\\
&& \widetilde{f}_2 = - \frac{1}{6 \, \GG^2} \quad \text{for} \quad
( \tilde{u}_1, \, \tilde{t} \, )
= \left( \frac{1}{\GG}, \, 0 \right) \,.
\end{eqnarray}
When $\GG > 2$,
the two solutions with nonvanishing $\tilde t$ disappear,
but we have all of the three critical points
(see~(\ref{rvalue})).
In the original set of variables $(u_1, \, t)$,
the critical points and the values of $f_2$ are given by
\begin{eqnarray}
& ( u_1, \, t ) \simeq ( 1.94928, \, \pm 0.966497 ) \,,
& f_2 \simeq -0.349830 \,,
\\
& ( u_1, \, t) \simeq ( 3.83475, \, 0 ) \,, \qquad \qquad
& f_2 \simeq -0.682622 \,.
\end{eqnarray}
We are interested in the critical points with a nonvanishing $t$,
which appear in pairs with opposite signs
due to the $\mathbb{Z}_2$ symmetry.
The value of $f_2$ at these critical points is approximately 35\%
of the predicted value.
It is encouraging that this coarse $n=1$ approximation gave
a potential which is of the same order
with the predicted value.  As we shall see, the result improves
quickly for $n>1$.

It is instructive to consider the effective twisted tachyon
potential as a function of the parameter~$\GG$.
The equation of motion for $\tilde{u}_1$ gives
\begin{equation}
\tilde{u}_1 = \frac{1 \mp \sqrt{1 - 4 \, \GG \, \tilde{t}^2}}
{2 \, \GG} \,.
\end{equation}
Correspondingly, there are two branches of the effective potential,
upper and lower:
\begin{equation}
\widetilde{f}_{\rm eff} =
\frac{-1 +6 \, \GG \, \tilde{t}^{\, 2}
-6 \, \GG^2 \, \tilde{t}^{\, 2}}
{12 \, \GG^2}
\pm \frac{(1 -4 \, \GG \, \tilde{t}^{\, 2} )^{\frac{3}{2}}}
{12 \, \GG^2} \,.
\end{equation}
The critical points are determined by the equation
\begin{equation}
\frac{\partial \widetilde{f}_{\rm eff}}{\partial \tilde{t}}
= \frac{\tilde{t}}{\GG}
\left( 1-\GG \mp \sqrt{1 - 4 \, \GG \, \tilde{t}^2} \, \right)
= 0 \,.
\end{equation}

\begin{figure}
\centerline{\epsfxsize=2in\epsfbox{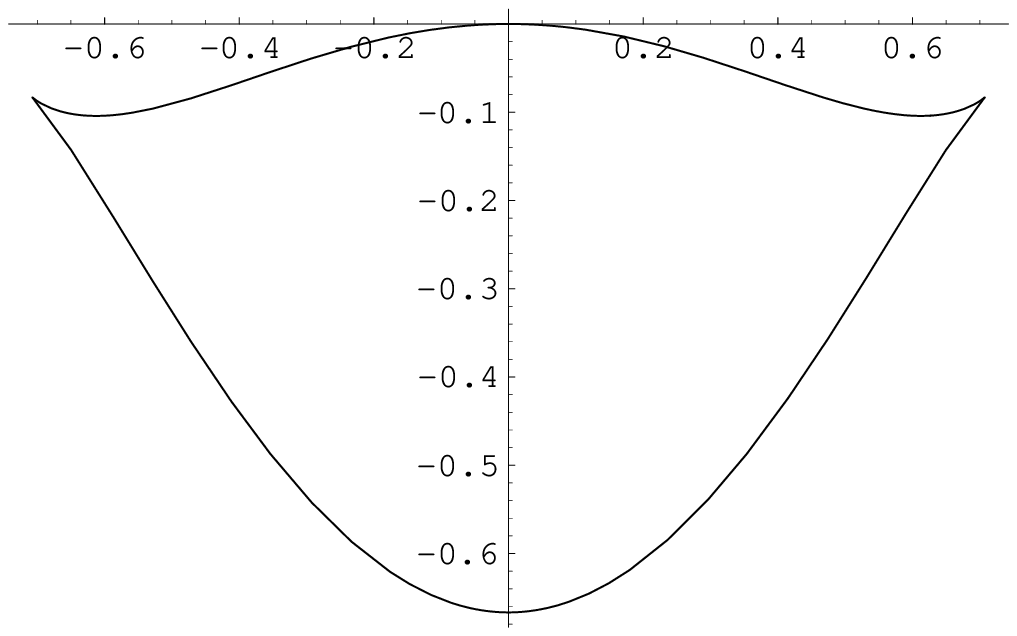}
\epsfxsize=2in\epsfbox{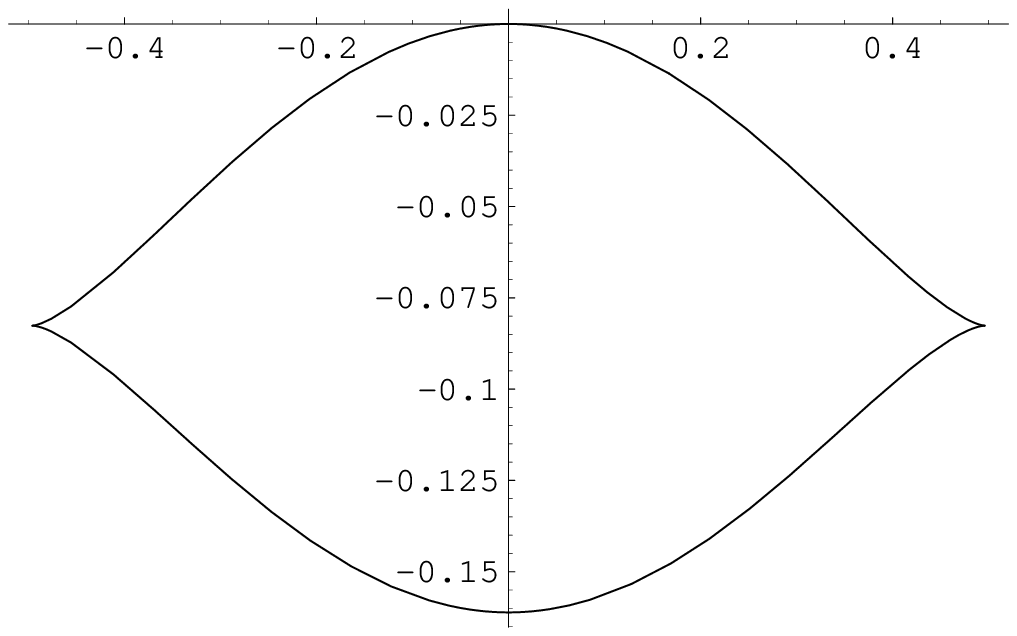}\epsfxsize=2in\epsfbox{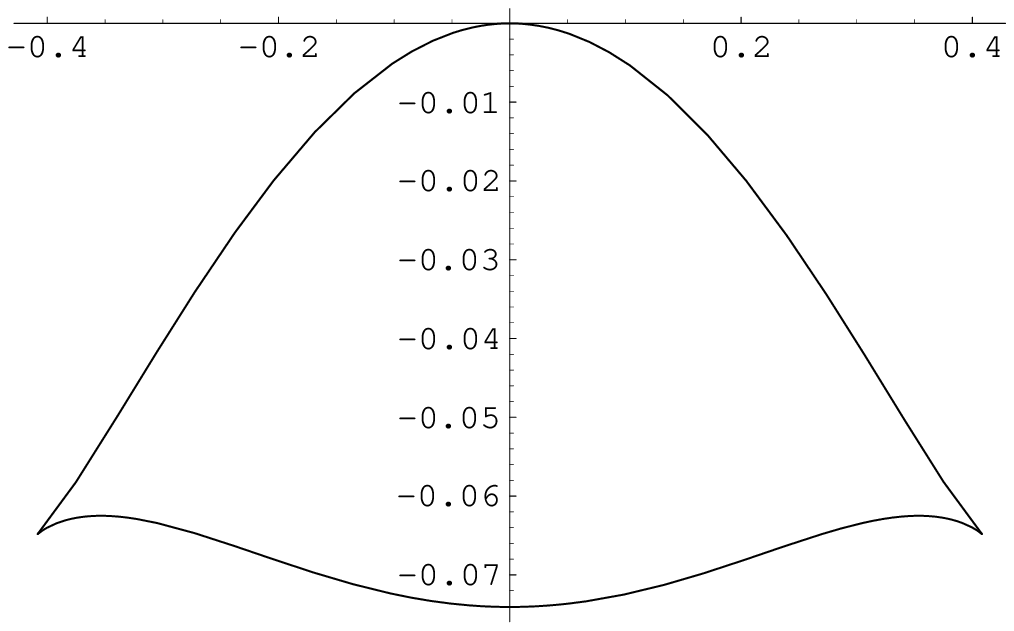}}
\caption{\small Effective potential
$\widetilde{f}_{\rm eff}(\tilde t)$
for $\GG=0.5$ (left), $\GG=1.017$ (center), and $\GG=1.5$ (right).
The two branches of the effective potential
meet at cusps located at $\tilde{t} = \pm 1/(2 \sqrt{\GG})$.
The critical points lie on the upper branch for $\GG=0.5$.
They pass  through the cusps at $\GG=1$
and move to the lower branch for $\GG > 1$.}
\label{eff}
\end{figure}

\noindent
The properties of $\widetilde{f}_{\rm eff}$ are
illustrated in Fig.~\ref{eff}.
For $0<\GG<1$ the nontrivial critical points lie
on the upper branch and are local minima.
For $1< \GG<2$ the critical points
lie on the lower branch and they are local maxima.
Despite appearances, nothing dramatic occurs
when $\GG$ goes through one.
The full potential (\ref{fullpotlev})
can be expanded about the critical points using
\begin{equation}
\tilde{u}_1 = \frac{1}{2} + \delta \tilde{u}_1 \,, \quad
\tilde{t} = \pm \frac{\sqrt{2-\GG}}{2} + \delta \tilde{t} \,,
\end{equation}
and the resulting quadratic terms are
\begin{eqnarray}
&& \frac{1}{2} \, (\GG-1) \, \delta \tilde{u}_1^2
\pm \sqrt{2-\GG} \, \delta \tilde{u}_1 \, \delta \tilde{t}
\nonumber \\
&=& \frac{1}{2} \left( \frac{1}{\sqrt{3-\GG}} \, \delta \tilde{u}_1
\pm \sqrt{\frac{2-\GG}{3-\GG}} \, \delta \tilde{t} \right)^2
- \frac{1}{2} \, (2-\GG)
\left( \mp \sqrt{\frac{2-\GG}{3-\GG}} \, \delta \tilde{u}_1
+ \frac{1}{\sqrt{3-\GG}} \, \delta \tilde{t} \right)^2 \,.
\qquad \quad
\end{eqnarray}
This quadratic form has one positive eigenvalue
and one negative eigenvalue
for the {\em entire} range $0 < \GG < 2$.
We interpret this as condensation of the twisted tachyon,
while the instability associated
with the bulk tachyon still remains.
We note that the mass eigenstates arise
through large mixing between
$\delta\tilde{u}_1$ and $\delta\tilde{t}$.
The effective potential only captures one direction
in the two-dimensional field space and the apparent mass
at the critical point does not bear
any simple relation to the mass eigenvalues
of the full potential.
In particular, the transition from $\GG < 1$ to $\GG > 1$
is completely regular.

\medskip
\noindent
$\underbar{Higher-order approximation}$.
For an arbitrary but fixed choice of $n$, $f_2$ can be written
as follows:
\begin{equation}
f_2 = - \frac{1}{2} \sum_{i=1}^n m_i^2 \, u_i^2
- \frac{1}{2} \, m_t^2 \, t^2
+ \sum_{i, \, j, \, k=1}^n g_{ijk} \, u_i \, u_j \, u_k
+ \sum_{i=1}^n c_i  \, u_i \, t^2 \,,
\end{equation}
where
\begin{eqnarray}
m_i^2 &=& \frac{1}{\pi} \left(
1 - \frac{\alpha_i^2}{\alpha_n^2} \, \xi_*^2 \right) \,,
\\
g_{ijk} &=& \frac{\xi_* \,
\RR^{6 -\frac{2 \, \xi_*^2}{\alpha_n^2}
(\alpha_i^2 + \alpha_j^2 + \alpha_k^2)}}
{3 \, \pi^{\frac{3}{2}} \, \alpha_n \,
J_1 (\alpha_i) \, J_1 (\alpha_j) \, J_1 (\alpha_k)}
\int_0^1 dy \, y \, J_0 (\alpha_i \, y) \, J_0 (\alpha_j \, y) \,
J_0 (\alpha_k \, y) \,,
\\
c_i &=& \frac{\alpha_n \, \RR^{\frac{11}{2}}}
{\pi^{\frac{3}{2}} \, \xi_* \, J_1 (\alpha_i) \,
\ln (\RR^2 \delta)}
\int_0^1 dy \, y \, e^{-\frac{\alpha_n^2}
{4 \, \xi_*^2 \, \ln (\RR^2 \delta)} \, y^2}
J_0 (\alpha_i \, y) \,.
\end{eqnarray}
We have carried out the numerical analysis
up to $n=30$.
As we increase $n$, we find a large number of critical points.
Most of them have a large negative value for $f_2$, which seems to
indicate that these solutions are not localized.
For each $n$, however, there are three critical points
on which $|f_2|$ remains small
and which correspond to the three critical points
we found for $n=1$.
The critical points with $t\not= 0$ appear as pairs with opposite
signs.  These critical points are equivalent and the quadratic
form of fluctuations about each of them has one positive
and $n$ negative eigenvalues.
This is consistent with our interpretation that
the twisted tachyon disappears,
while the instability associated with the bulk tachyon
remains.
For $t\not=0$ the values of $f_2$
obtained for $n=16, 17, \ldots, 30$ can be used to produce
the approximate fit:
\begin{equation}
\label{thecoorz2val}
f_2 = -0.720205 + \frac{1}{n} \, 0.522613 \,.
\end{equation}
This expression can be used to extrapolate $f_2$
in the limit $n \to \infty$
and gives $f_2 = -0.720205$.
Therefore, our level-truncation analysis gives approximately
72\% of the predicted value for the potential depth.

The third critical point has vanishing $t$.
We find a fit $f_2 = -2.23467 + 2.18465/n $.
This is a pure bulk tachyon solution whose physical interpretation
is not clear to us. We also do not know if this solution survives
the inclusion of the quartic term in the bulk tachyon potential.

\begin{figure}
\centerline{\epsfxsize=4in\epsfbox{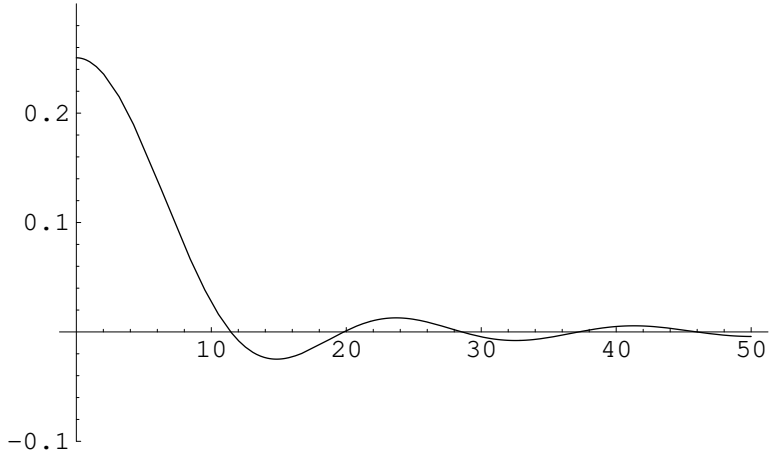}}
\caption{\small Coordinate-space profile $u(r)$ of the bulk tachyon
for $n=30$.
While $u(r)$ is required to vanish
at $r = \alpha_{30}/\xi_* \simeq 264.36$,
we see that it  is localized in a much shorter range.}
\label{localized}
\end{figure}

In  Fig.~\ref{localized} we show the coordinate-space
profile $u(r)$ of the bulk tachyon, calculated with $n=30$.
For this value of $n$ the cutoff radius of the cone, where
the field $u(r)$ is required to vanish, is $r\simeq 264.36$.
The figure illustrates that the bulk tachyon excitation is
well localized near the apex of the cone. The cutoff is
large enough to exhibit this localization and we have found that
the profile remains essentially unchanged  for larger $n$.

\medskip
It is natural to wonder if the level of the bulk tachyon
modes can be increased in a way consistent with a systematic
level expansion.  If we glance at Fig.~\ref{levexptabz2}, we see
that a new interaction appears at level $6/4$.
It may therefore make sense
to increase the level of the bulk tachyon field to
$5/12$ so that the $u^3$ interaction carries level less than or
equal to $5/4$.  We have repeated the analysis of the potential
for the corresponding new value of $\xi_*$ and find
\begin{equation}
\label{xiincreased}
f_2 = -0.503833 + {1\over n} \, 0.311575
\quad \text{for} \quad \xi_* = \sqrt{5/24} \,.
\end{equation}
The extrapolation thus gives  $f_2\simeq -0.504$,
which is worse than the result
obtained with lower momentum cutoff. Just as we saw in the
preliminary analysis of \S\ref{prelimanalesvlkd},
including higher level momentum modes of the bulk tachyon
makes the critical point shallower.
This result shows that the level expansion is not going
to provide a monotonic approach to the expected value of $f_2$.
It may be expected that as the level of interactions is increased
and new fields appear,
the value of $f_2$ will start to decrease.

\subsubsection{Momentum space method}

In this second approach, which we only sketch briefly,
we simply discretize the integrals
in the momentum space expression (\ref{f2momspace}) for $f_2$.
The discretization is characterized by an integer $N$.
We discretize $\xi = (\xi_1, \xi_2)$ as
\begin{equation}
(\xi_1, \xi_2)
= \Bigl(\, {i\over N} \, \xi_* \, , \,
{j\over N} \, \xi_* \Bigr),
\quad \text{with} \quad  -N \leq  i, j \leq N \,.
\end{equation}
The allowed discrete values of the momenta form
a square lattice. Since the smallest nonvanishing value of
$\xi_1$ (or $\xi_2$)
is $\xi_*/N$, the discretization
amounts to placing the system in a
two-dimensional square box of size $2\pi N/\xi_*$
and imposing periodic boundary conditions.

The integrals can then be written in terms of finite sums
that involve variables
$u[i,j] \equiv u[\xi_1, \xi_2]$.
The rotational symmetry of the problem allows us to reduce
the number of variables significantly.
Since $u[\,i,j] = u[\,i',j']$ whenever $i^2 + j^2 = i'^2 + j'^2$,
we can use a function of a single discrete variable
by setting $u[\,i,j]\to u[\,i^2+ j^2]$.
For a fixed $N$,
we further restrict the variables to
$u[\,i^2+ j^2]$ with $i^2 + j^2 \leq N$. In other words
we use disks in momentum space.  For $N=3$, for example, we
have seven variables
$u[0], u[1], u[2], u[4], u[5], u[8],$ and $u[9]$.

Integrals are easily discretized as follows:
\begin{equation}
\int {d^2\xi\over (2\pi)^2} \, F (\xi^2, u(\xi)) \simeq
\Bigl( {\xi_*\over 2\pi N } \Bigr)^2 \sum_{i^2 + j^2 \leq N^2}
\, F \Bigl( \,\, {\xi_*^2 \over N^2}(i^2 + j^2) \,, \,
u[i^2+ j^2] \Bigr) \,.
\end{equation}
The full expression for $f_2$, including multiple integrals,
is readily discretized and we have determined the critical points
of the potential for various values of $N$.
Taking $\xi_* = \sqrt{1/8}$,  and $N=3,4,5$, and $6$, we find
$f_2 = -0.710148, -0.735818, -0.706106$, and $-0.721683$,
respectively.
For higher values of $N$, the solution exhibits
decreasing oscillations about a value consistent
with (\ref{thecoorz2val}).
Taking $\xi_* = \sqrt{5/24}$ we find values fully consistent
with (\ref{xiincreased}).

\sectiono{The $\mathbb{C}/\mathbb{Z}_3$ tachyon potential}
\label{lec1.2}

The $\mathbb{C}/\mathbb{Z}_3$ orbifold has a tachyon potential
that is  more intricate than that of the $\mathbb{C}/\mathbb{Z}_2$
orbifold. First, the twisted tachyon is a complex field.
Second, pure cubic interactions of the twisted tachyon exist.
We find that the  tachyon potential has two inequivalent critical
points. In the deeper one, which we conjecture to represent
flat space, two tachyonic modes are lifted. In the shallower one,
which we conjecture to represent
the $\mathbb{C}/\mathbb{Z}_2$ orbifold, one tachyonic mode is lifted.

\subsection{Computing the action}

In addition to the untwisted sector with $k=0$,
the  $\mathbb{C}/\mathbb{Z}_3$ orbifold
includes two twisted sectors  $k=1$ and $k=2$.
The contributions to the potential from the untwisted sector
are the same as those of $\mathbb{C}/\mathbb{Z}_2$
and are given by (\ref{E_u^2+E_u^3}).
There are a pair of twist fields $\sigma_1$ and $\sigma_2$,
which we denote by $\sigma_1 = \sigma_{+} = \sigma_{--}$
and $\sigma_2 = \sigma_{-} = \sigma_{++}$ following
the conventions in \S\ref{section3}.
Their conformal dimensions are
$h_{\sigma_\pm} = \bar{h}_{\sigma_\pm} = 1/9$.
The mass-squared and the level
of the corresponding spacetime field $t$ are
\begin{equation}
m^2 = \frac{4}{\alpha'} \left( -\frac{8}{9} \right) \,, \quad
\ell(t) = \frac{2}{9} \,.
\end{equation}
The string field associated with the complex tachyon $t$
is written as
\begin{equation}
|T\rangle = \int {d^{D-2}q\over (2\pi)^{D-2}}
\Bigl( t(q) \, c_1 \bar c_1 |\sigma_+, q\rangle
+ t^*(-q) \, c_1 \bar c_1 |\sigma_-, q\rangle \Bigr) \,, \quad
t(q) = \int d^{D-2} x  \, t(x) \, e^{-iq\cdot x}\,.
\label{thestringfield}
\end{equation}
Since BPZ conjugation takes $|\sigma_{\pm}\rangle$
to $\langle \sigma_{\pm}|$ and Hermitian conjugation
takes $|\sigma_{\pm}\rangle$
to $\langle \sigma_{\mp}|$, this is a real string field
which leads to a real action.
The inner product is given by
\begin{equation}
\langle\sigma_\pm , -q' | \, c_{-1} \bar c_{-1} \,\,
c_0^- c_0^+ \,\, c_1 \bar c_1 | \sigma_\mp , q\rangle
= (2\pi)^{D-2} \, \delta^{(D-2)}(q+ q') \,,
\end{equation}
and the kinetic term for the twisted tachyon
and its contribution to the potential are found to be
\begin{equation}
- \int {d^{D-2}q\over (2\pi)^{D-2}} \, t^*(q)
\Bigl( q^2 - {4\over \alpha'}{8\over 9}\Bigr) t(q) \,, \quad
\mathbb{V}_{t^* t} = -\frac{4}{\alpha'}\frac{8}{9} \, t^* t \,.
\end{equation}

Cubic interactions of the twisted tachyon are allowed
by the $\mathbb{Z}_3$ symmetry.
The three-point function is given by
(\ref{answerthreepoint}) with $k/N=1/3$:
\begin{equation}
C_{+, +, +}
= \frac{3^{1/4}}{\sqrt{2 \pi^2 \alpha'}}
{\Gamma^2 (2/3) \over \Gamma (1/3)} \, V_{D-2}
= \frac{3^{3/4} \,
\Gamma^3 (2/3)}{2 \sqrt{2 \alpha'}\, \pi^2} \,V_{D-2}\,.
\end{equation}
The contribution to the potential from the cubic coupling
is given by
\begin{equation}
\mathbb{V}_{t^3+(t^*)^3}
= \frac{1}{3!} \, \frac{2 \kappa}{\alpha'} \,
\RR^{16/3} \, {2C_{+,+,+}\over V_{D-2}} \,
\Bigl(t^3 +(t^*)^3 \Bigr) \,,
\end{equation}
where the factor of $2$ in front of $C_{+,+,+}$ arises from the
ghost correlator. The coupling of bulk
and twisted tachyons is given by (\ref{cubutsquared}),
with an extra symmetry factor of two,
and the power of $\RR$  changed
due to the different conformal dimension:
\begin{equation}
\mathbb{V}_{u t^* t} = \frac{4 \kappa}{\alpha'} \,
\RR^{\frac{50}{9}} \, t^* t \int \frac{d^2 p}{(2 \pi)^2}
( \RR^2 \delta )^{-\frac{1}{4} \alpha' p^2} u(p) \,, \quad
\delta = 3^3 ~~ \text{for} ~~ \mathbb{C}/\mathbb{Z}_3 \,.
\end{equation}
We have now obtained all  the contributions to $\mathbb{V}_3$.
Using $\xi$ introduced in (\ref{scaledxip}),
setting
$u(\xi) = \frac{4 \kappa}{\alpha'} \, u(p)$,
and letting $t \to
\frac{\sqrt{\alpha'}}{\kappa} \, t$, the dimensionless
ratio $f_3$ can
be written as
\begin{eqnarray}
f_3 = \frac{3 \, \kappa^2 \, \mathbb{V}_3}{4 \pi}
&=& -\frac{8}{3 \pi} \, t^* t
-\frac{3}{8 \pi} \int \frac{d^2 \xi}{(2 \pi)^2} \,
u(-\xi) \left( 1 - \xi^2 \right) u(\xi)
\nonumber \\ &&
{}+ \frac{\beta}{4 \pi} \, \Bigl(t^3 +(t^*)^3 \Bigr)
+ \frac{3}{\pi} \, \RR^{\frac{50}{9}} \, t^* t
\int \frac{d^2 \xi}{(2 \pi)^2} \,
( \RR^2 \delta )^{-\xi^2} u(\xi)
\nonumber \\
&& {}+ \frac{1}{8 \pi}
\int \prod_{i=1}^{3} \left[
\frac{d^2 \xi_i}{(2 \pi)^2}\,\RR^{2(1-\xi_i^2)} \,
u(\xi_i) \right] (2\pi)^2 \delta^{(2)} (\xi_1+ \xi_2 + \xi_3) \,,
\end{eqnarray}
where
\begin{equation}
\beta = 2 \sqrt{\alpha'} \, \RR^{\frac{16}{3}} \,
{C_{+,+,+}\over V_{D-2}}
= \frac{3^\frac{3}{4} \, \RR^{\frac{16}{3}}}
{\sqrt{2} \, \pi^2}
\Gamma^3 (2/3)
\simeq 1.63674 \,.
\end{equation}
In coordinate space, $f_3$ is given by
\begin{eqnarray}
&& f_3 = \int_0^\infty dr \, r \,
\biggl( -\frac{3}{4} \, u(r) \left( \partial^2 + 1 \right) u(r)
+\frac{3 \, \RR^{\frac{50}{9}} \, t^* t}
{2 \pi \ln (\RR^2 \delta)} \,
e^{-\frac{r^2}{4 \ln (\RR^2 \delta)}} \, u(r)
+ \frac{1}{4} \left[ \RR^{2 (\partial^2+1)} u(r) \right]^3
\biggr)
\nonumber \\ && \qquad ~
{} -\frac{8}{3 \pi} \, t^* t
+ \frac{\beta}{4 \pi} \, \Bigl(t^3 +(t^*)^3 \Bigr) \,.
\end{eqnarray}
\begin{figure}[hbt]
\centerline{\epsfxsize=3.3in\epsfbox{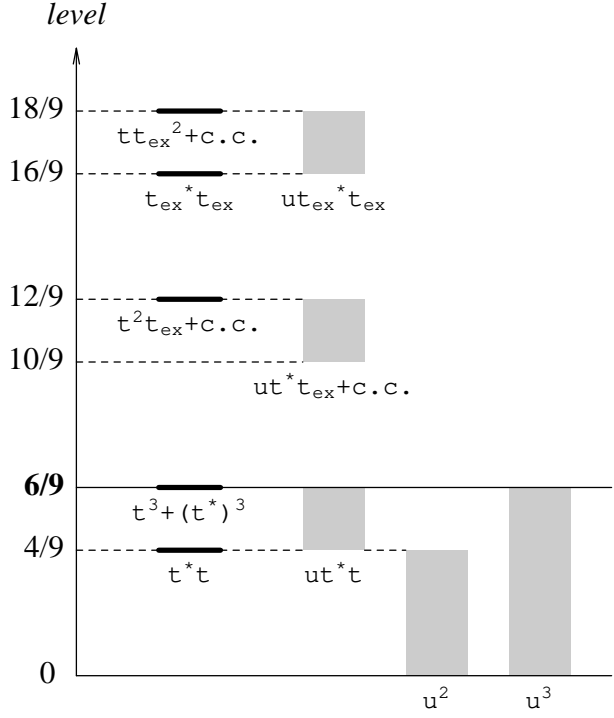}}
\caption{\small Level expansion table
for the $\mathbb{C}/\mathbb{Z}_3$ orbifold.
Since the bulk tachyon includes all levels from zero to $2/9$,
interactions that involve the bulk tachyon are shown as bands.
Terms involving excited twisted tachyons $t_{\rm{ex}}$ are shown,
but they are not included in our analysis.}
\label{levexptabz3}
\end{figure}
The next excited fields in the twisted sectors
have level $8/9$
and their kinetic terms have level $16/9$.
In our present analysis we will only include the twisted
tachyon $t$ of level $2/9$.  Accordingly, we restrict
bulk tachyon modes to have level less than or equal
to $2/9$.
This corresponds to the following momentum cutoff:
\begin{equation}
\ell(u(p)) = \frac{\alpha' p^2}{2}
= 2 \, \xi^2 \le \frac{2}{9}
\quad \to \quad
|\xi |\leq \xi_*= \frac{1}{3} \,.
\end{equation}
The resulting interactions are shown in Fig.~\ref{levexptabz3}.
Interactions that involve the excited twisted tachyons
are also shown in this figure,
but we do not include them in our analysis.

\subsection{A preliminary analysis}

\begin{figure}
\centerline{\epsfxsize=3in\epsfbox{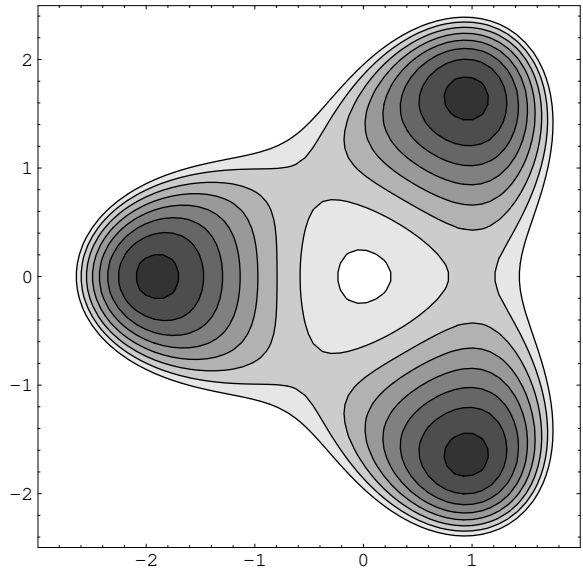}}
\caption{\small A contour-plot giving
the approximate effective potential for the complex tachyon $t$
of the $\mathbb{C}/\mathbb{Z}_3$ orbifold.
The horizontal and vertical axes correspond
to the real and imaginary parts of $t$, respectively.
The three saddle points and the three local minima respect
the $\mathbb{Z}_3$ symmetry.}
\label{figz3vac}
\end{figure}

Let us first study $f_3$ without the $u^3$ term
and the $t^3 + (t^*)^3$ term.
The equation of motion for $u(\xi)$ can be solved
in terms of $t^*t$,
and by substituting back into $f_3$,
we find the following effective potential:
\begin{equation}
\label{withoutt3}
f_3 = \frac{3}{4 \pi} \Bigl[ - m_{t^* t}^2 \, t^* t
+ \frac{\gamma}{2} \, (t^* t)^2 \Bigr] \,,\quad
m_{t^* t}^2 = \frac{32}{9} \,, \quad
\gamma = 16 \, \eta (\RR,\delta,\xi_*) \, \RR^{\frac{100}{9}}
\simeq 1.83463 \,.
\end{equation}
Since the $U(1)$ symmetry
$t \to e^{i \theta}\, t$ is unbroken, the potential has a minimum
at $|t| = m_{t^*t}/\sqrt{\gamma} \simeq 1.39213$ and
\begin{equation}
\label{miraculous}
f_3 = - {3m_{t^*t}^4\over 8 \pi \gamma} \simeq -0.822524 \,.
\end{equation}
This is a surprisingly good estimate.
The flat direction in the above potential is removed
when we introduce back the cubic terms $t^3 + (t^*)^3$:
\begin{equation}
\label{f3minusone}
f_3 = \frac{3}{4 \pi} \left[ - m_{t^* t}^2 \, t^* t
+ \frac{\beta}{3} \, \Bigl(t^3 +(t^*)^3 \Bigr)
+ \frac{\gamma}{2} \, (t^* t)^2 \right] \,.
\end{equation}
The $U(1)$ symmetry is broken to
$\mathbb{Z}_3$, and we find six critical points:
\begin{equation}
t_{\pm} = \frac{-\beta \pm \sqrt{\beta^2+4 \, \gamma \,
m_{t^* t}^2}} {2 \, \gamma} \,, \quad e^{2 \pi i/3} \, t_{\pm} \,,
~~\text{and}~~~ e^{4 \pi i/3} \, t_{\pm}\,.
\end{equation}
The critical points $(t_{-}, e^{2 \pi i/3} \, t_-, e^{4 \pi i/3}
\, t_-)$ are equivalent local minima of the effective potential,
while $(t_+, e^{2 \pi i/3} \, t_+, e^{4 \pi i/3} \, t_+)$
are equivalent saddle points (see Fig.~\ref{figz3vac}).
Substituting the values of the various parameters, we find
\begin{equation}
\label{valuescrit}
t_-\simeq -1.90792\,, ~f_3(t_-) \simeq -1.99722; \, \quad
t_+ \simeq 1.01578\,, ~ f_3(t_+) \simeq -0.369658 \,.
\end{equation}
While $f_3(t_-)$ changed considerably from the earlier
estimate in (\ref{miraculous}), it will turn out that
$f_3(t_-)$ comes back to a value near minus one
when we further include the $u^3$ terms.
This result, as well as the patterns of Fig.~\ref{levexptabz3},
suggests that the cubic terms in the twisted and bulk tachyons
should be included simultaneously.
We claim that the deeper critical point $t_{-}$
represents flat space.
The shallower critical point $t_+$ is a natural candidate
for the $\mathbb{C}/\mathbb{Z}_2$ orbifold.
As indicated in (\ref{z3predfac}),
we expect $f_3 = -0.25$
at the critical point that corresponds
to the $\mathbb{C}/\mathbb{Z}_2$ orbifold,
and the above result is in fact close to this value.
Further evidence for these identifications will arise
in the following subsection where we investigate the mass spectrum
around the critical points of the full potential.

\subsection{Level expansion analysis}
\label{section5.3}

Let us now study $f_3$ including the $u^3$ term.
It turns out that
the cubic interaction does not change the qualitative nature
of the critical points we obtained based on (\ref{f3minusone}).
We have carried out the level-truncation analysis numerically
using the coordinate space method in \S\ref{dlfkjgoin}.
As in the case of $\mathbb{C}/\mathbb{Z}_2$,
we found three branches of solutions
where $|f_3|$ remains small
as we increase $n$:
one with $t < 0$, one with $t > 0$, and one with $t=0$ .
The solution with a vanishing $t$ is not relevant
to our current interest.
The solutions with $t>0$ and  $t<0$
are not equivalent,
and they correspond to $t_{+}$ and $t_{-}$
in the previous subsection, respectively.
Using the data for $n=16, 17, \ldots, 30$,
the values of $f_3$ can be fitted as follows:
\begin{eqnarray}
\label{levtruncresz3}
f_3 &=& -0.988865 + \frac{1}{n} \, 0.886950 \quad
\text{for the branch with $t < 0$} \,,
\\
f_3 &=& -0.335624 + \frac{1}{n} \, 0.159658 \quad
\text{for the branch with $t > 0$} \,.
\label{thesecond}
\end{eqnarray}

For a given $n$, the full potential is a function of
$n+2$ real modes: $n$ from the bulk tachyon
and two from the twisted tachyon.
We have studied the mass spectrum around the critical points.
At the critical point with $t < 0$,
two modes become massive and $n$ modes remain tachyonic.
This is in fact the structure of flat space
with $n$ bulk tachyon modes included.
This provides further evidence that
the critical point with $t < 0$ represents flat space.
At the critical point with $t > 0$,
only one mode becomes massive and $n+1$ modes remain tachyonic.
This is in fact the tachyon structure
of the $\mathbb{C}/\mathbb{Z}_2$ orbifold
with $n$ bulk tachyon modes included.
It is therefore tempting to identify
the critical point with $t > 0$
as the $\mathbb{C}/\mathbb{Z}_2$ orbifold.

Our level-truncation result in (\ref{levtruncresz3})
gives about 99\% of the predicted value
for the depth of the critical point that represents flat space.
This accuracy is clearly accidental.
For the  critical point  argued to
represent the $\mathbb{C}/\mathbb{Z}_2$ orbifold,
the value $f_3 \simeq -0.34$ obtained from (\ref{thesecond})
is within reasonable range of the expected value $f_3= -0.25$.
It represents about 134\% of the expected depth.
In the same way as we discussed at the end of \S\ref{dlfkjgoin},
we can attempt to raise the level of the bulk tachyon modes
without conflicting with the rules of level expansion.
Assume we allow interactions with level up to $9/9$,
a value slightly below
the level of interactions involving new fields
in Fig.~\ref{levexptabz3}.
Then the maximum level allowed for tachyon modes is $1/3$, which
corresponds to $\xi_* = \sqrt{1/6}$.
For this cutoff value we find
\begin{eqnarray}
\label{levtrusdffdncresz3}
f_3 &=& - 0.691264 + {1\over n}\,  0.548867\,, \quad
\text{for the branch with $t < 0$} \,, \nonumber
\\
f_3 &=&  -0.277749 + {1\over n}\, 0.119452\,,  \quad
\text{for the branch with $t > 0$} \,.\nonumber
\label{thesecodfdfnd}
\end{eqnarray}
As in the case of $\mathbb{C}/\mathbb{Z}_2$,
the vacua have become shallower, but the values are still good.
It would be interesting to improve our analysis
by including higher-level fields and interactions.

\sectiono{Elementary four-point vertices}\label{elem4pt}

In closed string field theory
the quartic term $\langle \Psi, \Psi, \Psi, \Psi\rangle$ is an
integral of  a suitable measure over some
specific subspace $\mathcal{V}_{0,4}$ of the moduli space
$\mathcal{M}_{0,4}$ of four-punctured spheres.  Each
Riemann surface in $\mathcal{V}_{0,4}$ is nicely presented
by attaching a semi-infinite cylinder of circumference $2\pi$
to each of the faces of a tetrahedron.
The various surfaces arise
by varying the lengths of the edges of the tetrahedron,
subject to certain conditions.
Explicit computations require
the (very nontrivial) conformal map of such constructed surface
to the standard presentation as the Riemann sphere
$\widehat{\mathbb{C}}$ punctured at $0,1,x$ and $\infty$.
The (complex) value of $x$ depends on the parameters
of the tetrahedron. The images of the semi-infinite cylinders
divide $\widehat{\mathbb{C}}$ into four one-punctured disks.
For operators that are matter primaries times $c\bar c$,
the geometrical information needed to compute
the corresponding four-string vertex is
(i) the four mapping radii  $\rho_1, \rho_2, \rho_3$, and $\rho_4$
as a function of the modular parameter $x$ and
(ii) the region $\mathcal{D}$ such that
$x\in \mathcal{D}$ gives all the surfaces in $\mathcal{V}_{0,4}$.
Fortunately, Moeller~\cite{moeller} has been able to calculate
both the mapping radii and the domain $\mathcal{D}$ efficiently.
So, the numerical evaluation of any four-string interaction
that involves only primary operators is now possible.

The measure of integration for the quartic term
$\langle \Psi, \Psi, \Psi, \Psi\rangle$
in (\ref{action}) is given by a manifestly positive function
of the mapping radii
times the CFT correlator of the (matter) operators in question,
times a minus sign.
This minus sign is a nontrivial fact
about bosonic closed string field theory (the sign of
the CFT correlator is unambiguous because it is related
by factorization to products of three-point functions).
For bulk tachyons the CFT correlator is manifestly positive
and therefore the resulting quartic term in the potential
(minus the action) is negative definite
and tends to destabilize the vacuum.
Indeed, the elementary quartic term overwhelms the cubic term,
so the critical point of the cubic potential
disappears~\cite{Belopolsky:1994sk,Belopolsky:1994bj}.

It is natural to ask if the same fate awaits the potential
calculated in the present paper.  In fact, the CFT correlator
$Z(x, \bar x)$ of four twist fields is also manifestly positive,
so the quartic term of the twisted tachyons
will tend to destabilize the vacuum as well.
As we mentioned earlier,
we suspect that four-string interactions carry an intrinsic
level, perhaps of value near one.  If this is the case,
the inclusion of the various four-string elementary vertices will
typically be accompanied with the inclusion of other interactions
that involve excited tachyons of the twisted sector.
These latter interactions, just like those of the bulk tachyon,
tend to induce stabilizing terms that may possibly compensate
for the destabilizing effects
of the four-string interactions.
The full calculation should be done, but it is nontrivial.
Here we do a simple test:
we calculate the elementary quartic term of the twisted tachyon
and determine if it, alone, destabilizes
the vacuum found in the previous sections.
We find that it does not.

We consider, for definiteness,
the case of the $\mathbb{C}/\mathbb{Z}_3$ orbifold.
With the string field (\ref{thestringfield}),
the quartic contribution to the potential is
\begin{eqnarray}
\mathbb{V}_{(t^* t)^2}
&=&  {2\over \alpha'} \cdot {1\over 4!}\cdot
{\kappa^2\over V_{D-2}}
\langle
\Psi, \Psi, \Psi,
\Psi\rangle \nonumber\\
&=&  {2\over \alpha'} \, {1\over 4!} {i\over 2\pi}
\cdot  (2i) \cdot 2\cdot 6\,(t^* t)^2
\cdot {\kappa^2\over V_{D-2}} \int_{\mathcal{D}} du dv
{Z(x, \bar x) \over
(\rho_1 \rho_2\rho_3 \rho_4)^{2- {2\over 9}}}\,,
\end{eqnarray}
where $x = u + iv$.
The factor $(i/(2\pi))$ is part of the definition
of the four-string vertex
and the $2i$ arises from antighost insertions;
their product gives the sign mentioned above.
The factor of 2 is from the basic overlap
and the factor of 6 arises
because the product of four string fields contains
six terms with two $\sigma_+$'s and two $\sigma_-$'s.
As expected, each mapping radius appears
with the power of two minus twice the (matter) dimension
of the operator.
We then write
\begin{equation}
\mathbb{V}_{(t^*t)^2}
= -  {\kappa^2\over 2\alpha'} \,(t^*t)^2\,
{2\over \pi} \cdot
\int_{\mathcal{D}} {du dv\over
(\rho_1 \rho_2\rho_3 \rho_4)^2}
\,\,\Bigl[ {Z(x, \bar x)\over V_{D-2}}\cdot (\rho_1 \rho_2\rho_3
\rho_4)^{2\over 9}\, \Bigr]\,.
\end{equation}
Making use of (\ref{finreszxxbar}) we find
\begin{equation}
\label{thedeseriveds}
\mathbb{V}_{(t^*t)^2}= -  {\kappa^2\over {\alpha'}^2}
\,\frac{\sqrt{3}}{4\pi^2}\,(t^*t)^2\,\, {2\over \pi} \cdot
\int_{\mathcal{D}} {du dv\over
(\rho_1 \rho_2\rho_3 \rho_4)^2}
\,\,\Bigl[ \,\, \frac{|x(1-x)|^{-{4\over 9}}    (\rho_1
\rho_2\rho_3\rho_4)^{2\over 9}}
{F(x) F(1-\bar{x}) + F(1-x) F(\bar{x})}
\, \Bigr]\,.
\end{equation}
The integral
\begin{equation}
{2\over \pi} \cdot
\int_{\mathcal{D}} {du dv\over
(\rho_1 \rho_2\rho_3 \rho_4)^2} \simeq 72.39\,,
\end{equation}
was evaluated in order to obtain the quartic term in the bulk
tachyon potential~\cite{Belopolsky:1994bj, moeller}.
Interestingly, this information is sufficient to obtain
a reasonable estimate for the integral in (\ref{thedeseriveds}).
The expression inside brackets in (\ref{thedeseriveds})
turns out to be nearly constant
over the full region $\mathcal{D}$;
we found this by computing its value at special points
on $\mathcal{D}$ where the mapping radii are known.
The factor in brackets seems to vary from about 0.372
to about 0.390 (about a 5\% variation).
With the data provided by Moeller, we have actually found that
the bracket introduces an effective factor of 0.38002:
\begin{equation}
\mathbb{V}_{(t^*t)^2}
\simeq - {\kappa^2\over {\alpha'}^2}\,(t^*t)^2\,
\,\,\frac{\sqrt{3}}{4\pi^2}\, \,\cdot 72.39 \cdot 0.38002
\simeq -\,{\kappa^2\over {\alpha'}^2} \cdot 1.20693
\cdot (t^*t)^2 \,.
\end{equation}
The contribution to $f_3$ is approximately $-0.28813 \, (t^*t)^2$.
The critical points of the twisted
sector potential $\mathbb{V}_{t^* t} + \mathbb{V}_{t^3+(t^*)^3}$
in fact disappear when $\mathbb{V}_{(t^*t)^2}$ is included.
The critical points
of the full tachyon potential, however,
survive the inclusion of this destabilizing interaction term.
Indeed, repeating the analysis
in \S\ref{section5.3}
(for $n=30$) with this additional quartic term gives
$f_3\simeq -1.32$ for the vacuum that represents flat space
and $f_3 \simeq -0.71$ for the vacuum that
represents the $\mathbb{C}/\mathbb{Z}_2$ orbifold.
The elementary quartic interaction $\mathbb{V}_{(t^*t)^2}$
decreases $f_3$, as expected.
We have seen at the end of \S\ref{section5.3} that
higher-momentum modes of the bulk tachyon
increase $f_3$.
There are other interactions involving fields
at the next excited level which are expected to affect $f_3$
at the same order.
It would be interesting and important
to include these interactions in our analysis
to see if the critical points survive
and the values of $f_3$ improve.

\sectiono{Conclusions and discussion}

The computations we have done in this paper
provide new evidence for the conjecture
of Adams, Polchinski, and Silverstein~\cite{Adams:2001sv} (APS).
The potentials we obtain for low-level tachyons
in the $\mathbb{C}/\mathbb{Z}_2$
and $\mathbb{C}/\mathbb{Z}_3$ orbifolds
have the qualitative and quantitative features
required for tachyon condensation to trigger background changes
to orbifolds with smaller deficit angles or to flat space.
In our low-level calculations the metric is not dynamical
and the conjecture of Dabholkar~\cite{Dabholkar:2001if}
can be tested.   We find evidence that the tachyon potentials
have critical points at depths that roughly generate the expected
changes in deficit angles.  Our calculations were done
in the framework of closed string field theory, demonstrating
that this theory can be used to study nonperturbative
closed string physics.

We have found that the physically relevant potential
depends on twisted fields and localized bulk field excitations.
For computations that reach level two---the level of massless
closed string fields---the metric dependence
of $\mathbb{V}$ becomes explicit,
as we have to include the gravitational field into the
computations.  We have argued that for a metric dependent
potential the conjecture~\cite{Dabholkar:2001if} may not
hold exactly.
The metric dependence does not arise solely from the
localized bulk field excitations.
The Gaussian profile
of the CFT three-point function (\ref{theansatz})
shows that twisted fields are not $\delta$-functionally localized
but rather smeared out in a scale of order $\sqrt{\alpha'}$.
Their couplings to bulk fields thus contain form factors
that represent unusual metric dependence.
A calculation that includes gravity
would be of level $(2,6)$ and the typical level of our present
computations is $(1/4, 3/4)$.  We have some way to go before
we can discuss the effects of massless fields quantitatively.

In any computation beyond level two one must
include the  equation of motion for gravity.
Once the gravitational field is allowed to condense,
the APS conjecture cannot be tested
by looking at the change in the action, as we did in this paper.
Indeed, we saw in \S\ref{section2.1} that the action vanishes
at the classical solutions. It is not clear to us whether
there is a version of the Dabholkar conjecture that holds
once the gravitational back-reaction is included.
To find a way to test the APS conjecture
beyond level two is an important future problem.
One possibility may be to read off the deficit angle of the
spacetime from the metric that satisfies
the string field equations.
This may be challenging in the present form
of the string field theory action
where the direct identification of the graviton field with
the metric fluctuation is valid only for small fields.
Another possibility might entail the examination of boundary
contributions to the gravitational action.

\medskip
There are also a host of technical questions.
Additional work is necessary
in order to demonstrate that closed string field theory
can be used in some suitably defined level expansion.
There may also be other closed string backgrounds
that can be used to test
the level expansion \cite{Mukherji:1991tb,Sen:1993mh}.
The rules of closed string theory level expansion should be found.
The present calculations may be carried out to higher levels,
including systematically
the effects of (at least) quartic terms.
Since there are both positive and negative contributions
to the potential, it is important  to
verify that the critical points survive and
the ratios $f_N$ at the critical points
become closer to the predicted values.

It would be interesting to extend our
analysis of tachyonic instabilities to the
$\mathbb{R}/\mathbb{Z}_2$  orbifold,
which is not a cone but rather a half-line.
There is no prediction here for final state and
no conjecture concerning the critical points of the
tachyon potential.
It is also of interest to analyze orbifolds of finite volume,
in which case the end product of the decay
does not appear to be known either.
Finally, it would be very useful to develop
a workable closed superstring field theory of the NS-NS sector.
We could then test the original version of the conjectures
directly in the context of Type II backgrounds
that have no bulk tachyon.
It would be interesting to see
how the critical points
of the twisted tachyon potentials are generated.

\medskip
\noindent{\bf Acknowledgements}\\
\noindent
We would like to thank
A.~Adams, O.~Bergman, A.~Dabholkar, M.~Headrick, R.~Jackiw,
D.~Kutasov, H.~Liu, S.~Minwalla, N.~Moeller, J.~Raeymaekers,
M.~Schnabl, A.~Sen, W.~Taylor, and C.~Vafa
for useful discussions.
We are grateful to the authors of~\cite{Dabholkar} and~\cite{Adams}
for discussing their results prior to publication.
We also thank
N.~Moeller for providing the data necessary for the evaluation
of four-string interactions.
B.Z.~thanks the kind hospitality of Harvard University,
where part of this work was done.
This work was supported in part by
the DOE grant DF-FC02-94ER40818.

\end{document}